\documentclass[twoside]{article}
\usepackage{qic,epsfig}

\newcommand{\ms}[1]{\ensuremath{\mathsf{#1}}}
\newcommand{\bra}[1]{\ensuremath{\langle#1|}}
\newcommand{\ket}[1]{\ensuremath{|#1\rangle}}

\newcommand{\argmax}{\ensuremath{argmax}}

\newcommand{\ver}{{\ms{V}}}
\newcommand{\edge}{{\ms{E}}}
\newcommand{\nbr}{{\ms{nbr}}}

\newcommand{\energy}{{\mathcal{E}}}

\newcommand{\oy}{{\mathcal{Y}}}
\newcommand{\ham}{{\mathcal{H}}}

\newcommand{\tends}{\rightarrow}

\newcommand{\desev}{{\sc DeSEV}}

\newcommand{\gmin}{g_{\ms{min}}}
\newcommand{\wmis}{{\ms{mis}}}

\newcommand{\GCK}{G}

\newcommand{\art}{{\ms{ART}}}
\newcommand{\mat}{{\mathcal{M}}}

\def\final{1} % set this to 0 to get a comment-free version
\ifnum\final=1  %namely if we allow comments in the output
\newcommand{\vnote}[1]{[{\small Vicky: \bf #1}]\marginpar{*}}
\newcommand{\sidecomment}[1]{\marginpar{\tiny #1}}
\else % in this case [final=0] we don't want any comments to show
\newcommand{\vnote}[1]{}
\newcommand{\sidecomment}[1]{}
\fi  %ok, we're done with defining comment macros

\begin{document}
\setlength{\textheight}{8.0truein}    %FOR 2ND PAGE ONWARDS

%\runninghead{Different Adiabatic Quantum Optimization Algorithms  $\ldots$}
%            {V. Choi}

\normalsize\textlineskip
\thispagestyle{empty}
\setcounter{page}{1}

%\copyrightheading{Vol.}{No.}{Year}{Page Nos.}
%\copyrightheading{0}{0}{2011}{000--000}

\vspace*{0.88truein}

\alphfootnote

\fpage{1}

\centerline{\bf
%%%%%%%%%%%%%%%%%%%%%
%Put in titiles here
%%%%%%%%%%%%%%%%%%%%%
%DIFFERENT ADIABATIC QUANTUM OPTIMIZATION ALGORITHMS}
%\vspace*{0.035truein}
%\centerline{\bf FOR THE NP-COMPLETE EXACT COVER AND 3SAT PROBLEMS}
Avoid First Order Quantum Phase Transition by Changing Problem Hamiltonians}
\vspace*{0.37truein}
\centerline{\footnotesize
%%%%%%%%%%%%%%%%%%%%%%%%%%%%%%%%%%%%
%put authors' name and address here
%%%%%%%%%%%%%%%%%%%%%%%%%%%%%%%%%%%%
VICKY CHOI}
%FIRST AUTHOR\footnote{Typeset names in
%10 pt Times Roman, uppercase. Use the footnote to indicate the
%present or permanent address of the author.}}
\vspace*{0.015truein}
\centerline{\footnotesize\it Department of Computer Science, Virginia
  Tech, 7054 Haycock Road}
\baselineskip=10pt
\centerline{\footnotesize\it Falls Church, VA 22043, USA}

%\publisher{(received date)}{(revised date)}

\vspace*{0.21truein}

%% \abstracts{first paragraph}{second paragraph}{third paragraph}
%% If there is only one paragraph, just keep the second and third empty 
%% like the following one 
\abstracts{
%%%%%%%%%%%%%%%%%%%%
% put abstract here
%%%%%%%%%%%%%%%%%%%%
In Amin and Choi~\cite{AC09}, we show that an adiabatic quantum
algorithm for the NP-hard maximum independent set (MIS) problem on a set of special
family of graphs in which there are exponentially many local maxima
would have the exponentially small minimum spectral gap and 
thus would require the exponential time, due to the first order quantum
phase transition (FQPT).
The problem Hamiltonian of the adiabatic quantum algorithm for MIS 
is based on the reduction to the Ising problem and has flexible
parameters. 
In this paper, we show numerically on the 15-vertex graph that 
by choosing the parameters appropriately in the problem Hamiltonian (without changing the problem to be solved)
for MIS, we can prevent the FQPT and drastically increase the minimum spectral gap.
The result is further supported by visualization from the Decomposed
State Evolution Visualization (\desev) --- a visualization tool we introduced.
Furthermore, our result also serves to
concretely
clarify that
it is not sufficient to consider one specific problem Hamiltonian
for
proving the failure of adiabatic quantum optimization for a problem, as
explained in \cite{ChoiDiff}.
We also raise the basic question about what the appropriate formulation of
adiabatic running time should be. 
}{}{}

\vspace*{10pt}

%\keywords{Adiabatic Quantum Optimization, Adiabatic Quantum
%  Algorithms, NP-Complete, Exact Cover, 3SAT, Maximum Independent Set}
%\vspace*{3pt}
%\communicate{to be filled by the Editorial}

\vspace*{1pt}\textlineskip    %) USE THIS MEASUREMENT WHEN THERE IS
   %) A SECTION HEADING
%\vspace*{-0.5pt}
%\noindent
%%%%%%%%%%%%%%%%%%%%%%%%%%%%%%%%
%put the text of the paper here
%%%%%%%%%%%%%%%%%%%%%%%%%%%%%%%%
\section{Introduction}

Adiabatic quantum computation (AQC) was proposed by
Farhi~et~al.~\cite{FGGS00,FGGLLP01} in 2000 as an alternative quantum paradigm to solve NP-hard 
optimization problems, which are believed to be classically
intractable.
The same idea to the adiabatic quantum optimization,
under a different name of
 {\em quantum annealing},  was first put forward by Apolloni et al. in
 1988, 
 see \cite{ST2006,ST2008} and references therein for a history of
 the field.
It was shown that AQC is not
just limited to optimization problems, and is polynomially equivalent to 
 conventional quantum computation (quantum circuit model) \cite{ADKLLR04,lidar-equiv}. 
A quantum computer promises extraordinary power over a classical computer,  as demonstrated by 
Shor~\cite{shor} in 1994 with the polynomial quantum algorithm for solving the factoring problem, 
for which the best known classical algorithms are exponential.
Just how much more powerful are quantum computers? 
In particular, we are interested in whether 
an adiabatic quantum computer can
 solve NP-complete problems  
{\em more} efficiently than a classical computer. 

Unlike classical computation or  quantum circuit model 
in which an algorithm is specified by a 
finite sequence of {\em discrete} operations via classical/quantum gates, the adiabatic 
quantum algorithm is {\em continuous}.
It has been assumed (see Section~\ref{sec:AQA} for more discussion) that, 
according to the adiabatic theorem, 
the dominant factor of the adiabatic running time (\art) of the algorithm 
scales polynomially with the inverse
of the {\em minimum spectral gap} $\gmin$ of the system Hamiltonian (that describes the algorithm). 
%see Section~\ref{sec:ART} for more discussion on the adiabatic running time.
Therefore, in order to 
 analyze the running time of
an adiabatic algorithm, it is necessary to be able to 
bound $\gmin$ analytically. 
However,  $\gmin$ is in general difficult to compute (it is as
hard as solving the original problem if computed directly).
Rigorous analytical analysis of adiabatic algorithms remains challenging.
%\vnote{There are no rigorous studies about the performance...}
Most studies have to 
resort to numerical calculations.
These include
numerical integration of Schr\"odinger equation
\cite{FGGLLP01,childs-clique}, %vc: Farhi et al did not compute the gaps.
eigenvalue computation (or exact diagonization)\cite{znidaric-2005-71,symmetries}, and quantum Monte Carlo (QMC) technique
\cite{Young,young-2009}. 
However,  not only are these methods  limited to small sizes 
(as the simulations of
quantum systems grow exponentially with the system size), but also little insight can be gained from these  numbers
to design and analyze the time complexity of the algorithm.
%alone, e.g., how does the minimum spectral gap relate to the problem structure?

Perhaps, from the algorithmic design point of view, it is more 
important to unveil the quantum evolution black-box
and thus enable us to obtain insight for designing
efficient  adiabatic quantum algorithms.
For this purpose, we devise a visualization tool, called Decomposed
State Evolution Visualization (\desev).

One of the original ideas of AQC in \cite{FGGS00,FGGLLP01} was proposed 
 as an energy minimization algorithm
 that aims to use coherent quantum evolution to avoid trapping 
in the local minima 
that trip classical algorithms of NP-hard optimization problems.
However, several works \cite{DV01,DMV01,Reichardt-04} showed that
their adiabatic quantum 
 algorithm failed to avoid local minima and would take exponential
 time for some problem. As we pointed out in \cite{ChoiDiff}, their
 lower bounds are only for their specific adiabatic quantum algorithms,
 and their arguments are not sufficient for showing the lower bound of
 all adiabatic quantum algorithms of the same problem. Nevertheless,
 one still might argue
 that their results provide
``convincing evidence'' that AQC would fail to solve problems with
many local minima. 
For this purpose, we set out to study graphs that have exponential
many local maxima for the NP-hard Maximum
Independent Set (MIS) problem which AQC ``naturally'' solves, see
Section~\ref{sec:MIS} and Choi~\cite{minor-embedding}. While it is not difficult to
come up with such graphs, the challenge lies in coming up such graphs
with small sizes such that we can visualize the quantum evolution
using \desev, which relies on the (numerical) eigenvalue computation.
After many trial-and-errors, with the aid of \desev,  
we  constructed a special family of graphs in which there are
exponentially many local maxima for MIS, with the smallest size of
such graph being $15$. For the reference sake, we call this family of
graphs the CK graph.
The numerical results of an adiabatic algorithm for MIS on these graphs suggested
that $\gmin$ is exponentially small and thus the algorithm requires exponential time.  These results
were then explained by the first order quantum phase transition (FQPT)
by Amin and Choi in \cite{AC09}. 
That is, our result agreed with the speculation --- the system got trapped in the
local minima  and the particular AQO algorithm failed due to the FQPT.
Since then, there have been some other papers (Altshuler
et al.,~\cite{altshuler-2009} ; Farhi et al., ~\cite{farhi-2009}; Young et al.,~\cite{young-2009}; Jorg et al.,~\cite{Jorg1,Jorg2})
 investigating the same phenomenon, i.e., first order
quantum phase transition.
% on \EC and 3SAT. \vnote{there are some more references...}
In particular, Farhi~et~al. in~\cite{farhi-2009} suggested that the exponential small gap caused by the FQPT
 could be overcome (for the set of instances they consider) by randomizing the choice of initial Hamiltonian. 
In this paper, 
we show numerically that 
by 
changing the parameters in the problem Hamiltonian (without changing the problem to be solved) of the adiabatic algorithm for 
MIS on the CK graph, we prevent the FQPT from occurring and significantly
increase $\gmin$. 
We further support our result by visualization from \desev.
To the best of our knowledge, this is the first time that such a
numerical result has been shown.
We do so by scaling the vertex-weight of the graph, namely, multiplying the weights of vertices by a scaling factor.
In order to determine the best scaling factor, we raise the basic
question about what the appropriate formulation of adiabatic running
time should be. 
We remark that the scaling factor in turn relates to the bit of
precision required for the parameters in the Hamiltonian which is an
important physical resource and plays a critical role in the
computational complexity.
Finally, our result serves to further clarify (see \cite{ChoiDiff,Choi-PNAS}
for explanation) that 
it is not sufficient to consider one specific problem Hamiltonian
for
proving the
the failure of adiabatic quantum optimization for a problem.
% as in
%\cite{DV01,altshuler-2009}.

%\vnote{revising needed}

This paper is organized as follows. 
In Section~\ref{sec:AQA}, we review the adiabatic quantum algorithm, and
the adiabatic running time(\art). 
In Section~\ref{sec:MIS}, we recall the adiabatic quantum algorithm for MIS based on the reduction to the
Ising problem.
In Section~\ref{sec:desev}, we describe the visualization tool \desev{} and the CK graph. We show examples of 
\desev{} on the MIS adiabatic algorithm for the CK graph. 
In Section~\ref{sec:change-parameter}, we describe how changing the parameters affects $\gmin$, and raise the question about $\art$.
We conclude with the discussion in Section~\ref{sec:discussion}.

\section{Adiabatic Quantum Algorithm}
\label{sec:AQA}
An adiabatic quantum algorithm is described by a time-dependent system Hamiltonian
\begin{equation}
\ham(t) = (1-s(t))\ham_{\ms{init}} + s(t) \ham_{\ms{problem}}
\end{equation}
for $t \in [0,T]$, $s(0)=0$, $s(T)=1$.
There are three components of $\ham(.)$: 
(1) initial Hamiltonian: $\ham(0)=\ham_{\ms{init}}$;
(2) problem Hamiltonian:  $\ham(T)=\ham_{\ms{problem}}$;
and (3) evolution path: $s : [0,T] \longrightarrow [0,1]$, e.g., $s(t)=\frac{t}{T}$.
$\ham(t)$ is an adiabatic algorithm for an optimization problem if we encode the problem into the problem 
Hamiltonian $\ham_{\ms{problem}}$ such that the ground state of $\ham_{\ms{problem}}$ corresponds to the answer to
the problem.

In this paper,  we fix the evolution path by the linear interpolation function $s(t)=\frac{t}{T}$. 
Hereafter,  we describe an adiabatic algorithm by the re-parametrized  Hamiltonian
\begin{equation}
\ham(s) = (1-s)\ham_{\ms{init}} + s \ham_{\ms{problem}}
\end{equation}
where $s \in [0,1]$, with understanding that $s(t)=t/T$.
For a more general interpolation path see \cite{lidar-path}.
Furthermore, throughout this paper, we fix the initial Hamiltonian to be  $\ham_{\ms{init}} = - \sum_{i \in \ver(G)}  \sigma_i^x$.
%In other words, different adiabatic algorithms for the same problem are described by different problem Hamiltonians for the problem. 
When it is clear from context, we also refer to the problem Hamiltonian as the adiabatic algorithm for the problem.
%Therefore, we refer the different adiabatic algorithms for a problem simply by the different problem Hamiltonian.
%Thus, we refer to the problem Hamiltonian as the adiabatic algorithm for the problem.
%\vnote{revise this sentence?}

\medskip
\subsection{Adiabatic Running Time}
\label{sec:ART}

In their original
work~\cite{FGGS00}, the running time of the adiabatic
algorithm is defined to be the same as the adiabatic evolution time
$T$, which is given by the adiabatic condition of 
the adiabatic theorem. 
However, this definition is under the assumption of some physical limit of the maximum energy of the system (see e.g.,~\cite{jordan1}),
and is not well-defined from the computational point of view, as observed by
Aharonov~et~al.~\cite{ADKLLR04}. 
They re-define $\art(\ham)$ as 
$T \cdot
\max_s||\ham(s)||$, taking into the account of the time-energy 
trade-off in the Schr\"odinger's equation\footnote{ 
Namely,
$
i\frac{d\ket{\psi(s)}}{ds} = T\cdot \ham(s) \ket{\psi(s)}
= \frac{T}{K}\cdot K \ham(s) \ket{\psi(s)} 
$
where $K>0$ is a constant.}.

%However, it is interesting (if not confusing) that many different versions of the
%adiabatic condition(s) of the adiabatic theorem
%have been recently proposed.
On the other hand, given the extensive work on the rigorous proofs of the adiabatic theorem, 
it is interesting (if not confusing)  that many different versions of the
adiabatic conditions
have been recently proposed. 
These include 
 \cite{adt-1,adt-2,adt-3,adt-4,adt-5,adt-6,adt-7,adt-8,adt-9,adt-10,adt-11} in
    the quantum physics community, and
\cite{Reichardt-04,ADKLLR04,adt-AR} in
the computer science community.
%Under the assumption that the Hamiltonian changes with the size of
%the problem is at most in the order of polynomial, 
%all these studies agree that for large problem size $n$, 
Most of these studies imply that
\art{} scales polynomially with the inverse of the spectral gap
of the system Hamiltonian, which is 
%Therefore, the current formulation of \art{} is
sufficient when one is interested in the coarse computational complexity
of algorithms, namely, the distinction between polynomial and
exponential running time.
%\vnote{reviewer complaints about `suggest', perhaps show}
%\vnote{Is this still correct?}

However, from both the practical and algorithmic point of view, 
it is important to have a more precise formulation of \art. 
First, this is because the specification of the adiabatic evolution
time $T$ is required in an adiabatic algorithm, and
therefore a tight and simple upper bound is desired.
That is, while there are complicated formulas such as the ones from \cite{adt-11}, although accurate, they
are not useful if the formulas can not be efficiently evaluated.
%\vnote{This is in contrast to the complicated formulas which although
 % accurate, it is not useful if the formula can not be efficiently computed.} 
Second, we are interested in the actual time complexity of the
algorithm, and not just the polynomial vs. exponential distinction.
It is necessary to have a more precise formulation of \art{} such that
basic algorithmic analysis can be carried out. 
Third, at this stage of research, 
it is particularly important to have such a formulation
because the spectral gap, which plays the dominating role in the
formulation of \art, is difficult to analyze. All current efforts
on the spectral gap analysis
%(see Section \ref{sec:spectral-gap})
resort to numerical studies, 
and that means the studies are
 restricted to small problem sizes only. Therefore, to gain 
insight into the time complexity of algorithms from these small
instances, it is important that the formulation of \art{} applies to small sizes.
So what is the appropriate formulation of \art? What should the adiabatic condition(s) be? 
This is in contrast to the study in ~\cite{lidar-path} where the exact form of adiabatic condition is not essential.
%And is there a universal
%condition? Or are there different sets of conditions for different classes
%of adiabatic algorithms?
%Readers are referred to \cite{ART-paper} for some discussion. 
In Section~\ref{sec:art}, we compare three closely related versions and 
raise the question about what the appropriate adiabatic running time should be.

\section{An Adiabatic Algorithm for MIS}
\label{sec:MIS}
In this section, we recall the adiabatic algorithm for MIS that is based on the reduction to the Ising problem, as described in \cite{minor-embedding}. 
First, we formally define the Maximum-Weight Independent Set (MIS)
problem (optimization version):

\smallskip
\hspace*{0.7cm}{\bf Input:} An undirected graph $G =(\ver(G),\edge(G))$, where each vertex $i \in \ver(G) = \{1, \ldots, n \}$ is weighted by a
positive rational number $c_i$

\hspace*{0.7cm}{\bf Output:} A subset $S \subseteq \ver(G)$ such that
$S$ is independent (i.e., for each $i,j \in \ver(G)$, $i\neq j$, $ij
\not \in \edge(G)$) and the total
{\em weight} of $S$ $=\sum_{i \in S}
c_i$ is maximized. 
Denote the optimal set by $\wmis(G)$.
\smallskip

%\paragraph{Notation.} We denote the instance by $<\!G,\{c_i\}\!>$, and the optimal set by $\wmis(G)$.
%Note that the integer-weighted {\sc MIS} (and hence the positive
 %rational weighted one) is polynomially
 %reducible to the unweighted {\sc MIS}.
%~\cite{wvc-paper}.

There is a one-one correspondence between the  MIS problem and the Ising
problem, which is the problem directly solved by the quantum processor
that implements 1/2-spin Ising Hamiltonian. We recall the
quadratic binary optimization formulation of the problem.
More details can be found in \cite{minor-embedding}.
\begin{theorem}[Theorem 5.1 in \cite{minor-embedding}]
If $J_{ij} \ge \min\{c_i,c_j\}$ for all $ij \in \edge(G)$, then the maximum
  value of
  \begin{equation}
\oy(x_1,\ldots, x_n) = \sum_{i \in \ver(G)}c_i x_i - \sum_{ij \in \edge(G)}
  J_{ij}x_ix_j
\label{eq:Y}
  \end{equation}
is the total weight of the MIS. 
In particular if $J_{ij} > \min\{c_i,c_j\}$ for all
      $ij \in \edge(G)$, then $\wmis(G) = \{i \in \ver(G) : x^*_i = 1\}$,
where $(x^*_1, \ldots, x^*_n) = \argmax_{(x_1, \ldots, x_n) \in \{0,1\}^n}
\oy(x_1, \ldots, x_n)$.
\label{thm:mis}
\end{theorem}

Here the function $\oy$ is called the pseudo-boolean function for MIS.
Notice that in this formulation, we only require $J_{ij} > \min\{c_i,c_j\}$, and thus there is freedom in
choosing this parameter. In this paper we will show how to take advantage of this.

By changing the variables ($x_i=\frac{1+s_i}{2}$), it is easy to show that MIS is equivalent
to minimizing the following function, known as the {\em Ising energy function}:
\begin{eqnarray}
  \energy(s_1, \ldots, s_n) &=& \sum_{i \in \ver(G)} h_i s_i + \sum_{ij \in \edge(G)} J_{ij}s_is_j,
\end{eqnarray}
which is the 
eigenfunction of the following
 {\em Ising Hamiltonian}:
%which is the energy function (eigenvalues) of 
%{\em Ising
%Hamiltonian}:
%\footnote{That is, replace each variable $s_i \in \{-1,+1\}$ by $\sigma^z_i$.}:
\begin{equation}
\ham_{\ms{Ising}} = \sum_{i \in \ver(G)} h_i \sigma^z_i + \sum_{ij \in \edge(G)} J_{ij}
\sigma^z_i \sigma^z_j
\label{eq:Ising}
\end{equation}
where $h_i = \sum_{j \in \nbr(i)}
  J_{ij} - 2c_i$, $\nbr(i) =\{j: ij \in \edge(G)\}$,
for $i \in \ver(G)$.
%The Ising energy function is the 
%where $h_i = \sum_{j \in \nbr(i)}
%  J_{ij} - 2c_i$, for $i \in \ver(G)$.

% That is, an adiabatic algorithm for MIS in which the problem Hamiltonian is $\ham_{\ms{Ising}}$ 
% is described by the following 
% system Hamiltonian:
% \begin{equation}
% \ham(s) = (1-s)\ham_{\ms{init}} + s \ham_{\ms{Ising}}
% \end{equation}
%  where $s \in [0,1]$ with the assumption that $s(t)=t/T$.
% If $T$ is sufficiently large according to the adiabatic theorem, then 
% the ground state of $\ham(1)$, say $\ket{x_1^*x_2^*\ldots x_n^*}$,
%   corresponds to the maximum-weight independent set, namely $\wmis(G) =
%   \{i: x_i^* = 0\}$\footnote{Notice we use $x_i=\frac{1+s_i}{2}$ instead of $x_i=\frac{1-s_i}{2}$.}.

\section{\desev{} and CK Graph}
\label{sec:desev}
In this section, we describe a visualization tool, called Decomposed State Evolution Visualization (\desev), which aims to 
``open up''  the quantum evolution black-box from a computational point of view. 
Consider the above adiabatic algorithm for MIS. 
Recall that according to the adiabatic theorem, if the evolution is slow enough, the system remains in the instantaneous ground state.
Let $\ket{\psi(s)}$ be the ground state of $\ham(s)$, for $s \in [0,1]$. For a system of $n$-qubits, $\ket{\psi(s)}$ is a superposition
of $2^n$ possible computational states, namely,
$$\ket{\psi(s)} = \sum_{x \in \{0,1\}^n} \alpha_x(s)
\ket{x}, \mbox{         where }\sum_{x \in
  \{0,1\}^n}|\alpha_x(s)|^2 =1.$$
For example, we have the initial ground state $\ket{\psi(0)} =
\frac{1}{\sqrt{2^n}} \sum_{x \in \{0,1\}^n}\ket{x}$, which is the uniform superposition of all $2^n$ states, while the final ground state $\ket{\psi(1)}= \ket{x_1^*x_2^*\ldots x_n^*}$, corresponding to the solution state.
A natural question is: what are the instantaneous ground states $\ket{\psi(s)}$, for $0<s<1$, like? In particular, we would like to ``see''
how the instantaneous ground state evolves? 
A na\"{i}ve solution would be to trace the $2^n$ amplitudes $\alpha_x$.
% change, but there are $2^n$ of them. 
The task becomes unmanageable even for $n=10$, which has $1024$ amplitudes, even though many may be negligible (close to zero). 
%there are still too many numbers to trace.

To make the ``visualization''  feasible, we introduce a new measure $\Gamma_k$.
Suppose that $\ham(1)$, has $(m+1) \le 2^n$
distinct energy levels: $E_0 < E_1 < \ldots < E_m$.
For $0 \le k \le m$, 
let $D_k= \{x \in \{0,1\}^n: \ham(1)\ket{x} = E_k \ket{x}\}$ be the set of (degenerate)  computational states that have the same energy level $E_k$ (with respect to  the problem Hamiltonian $\ham(1)$), and
define 
$$ \Gamma_k(s) = \sum_{x \in D_k} |\alpha_x(s)|^2.$$
In other words, $\Gamma_k(s)$ is the total percentage of (computational) 
states of the same energy level $E_k$ participating in $\ket{\psi(s)}$. 
 The idea is now to trace $\Gamma_k$  instead of $\alpha_x$. 
Here we remark that  $\Gamma_k$ are defined for any eigenstate $\ket{\psi}$ and not just for the ground state.

% For this method to work, we need to carefully 

For our purpose, 
% To study the evolution,
we constructed a special
family of vertex-weighted graphs for the MIS problem. 
We designed the problem instances such that the global minimum is ``hidden'' in the sense that 
 there are many local minima to mislead local search based algorithms. 
%many 
%states are of the same energy level (which serve as local minima), and thus there are a lot less (than $2^n$) energy level.
Note that  the size of the smallest  instances needs to be necessarily
smaller than $20$ as we are relying on the eigenvalue computation (or
exact diagonization) to
compute $\Gamma_k$.

\begin{figure}[h]
$$
  \begin{array}{cc}
\includegraphics[width=0.5\textwidth]{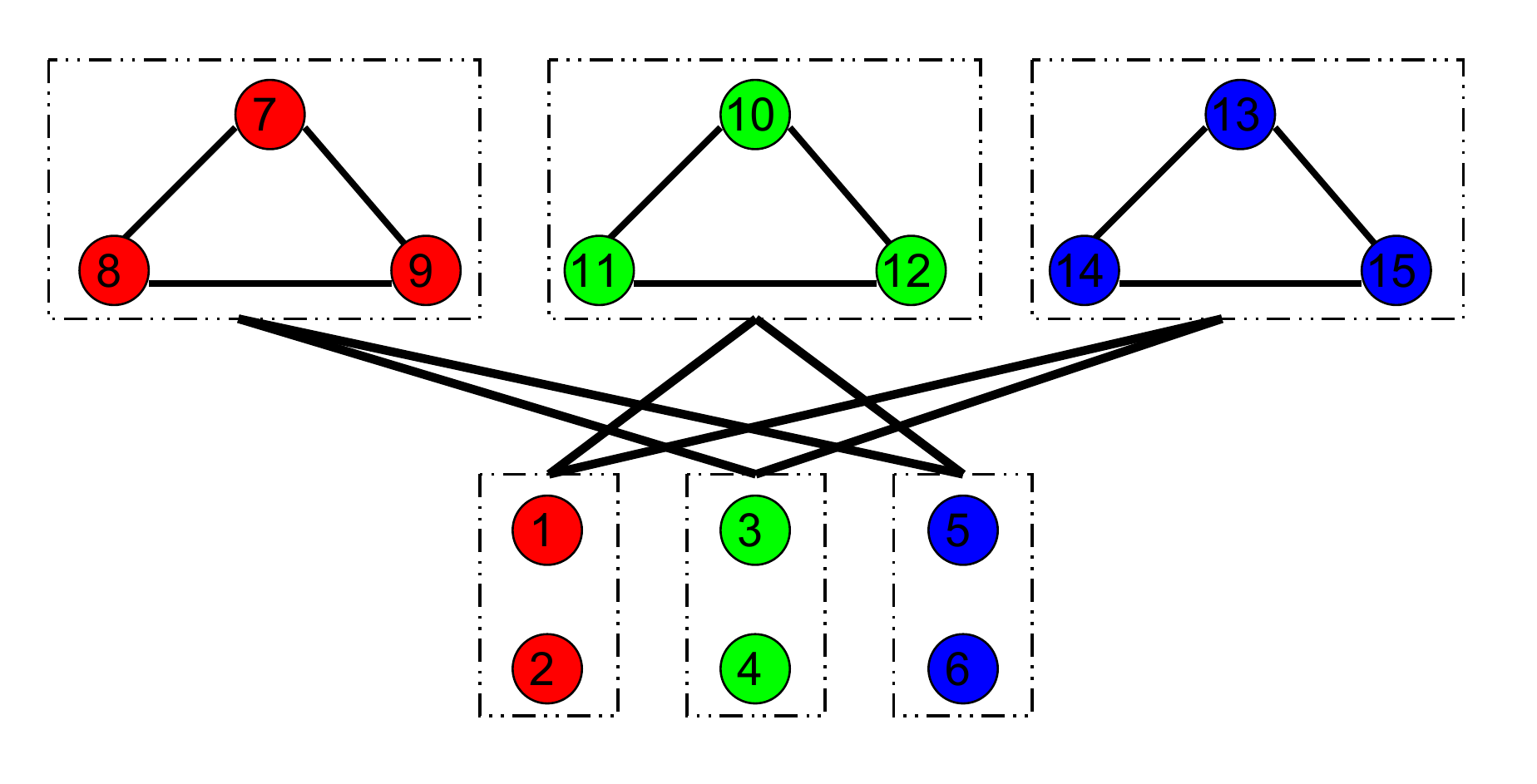} &
\includegraphics[width=0.3\textwidth]{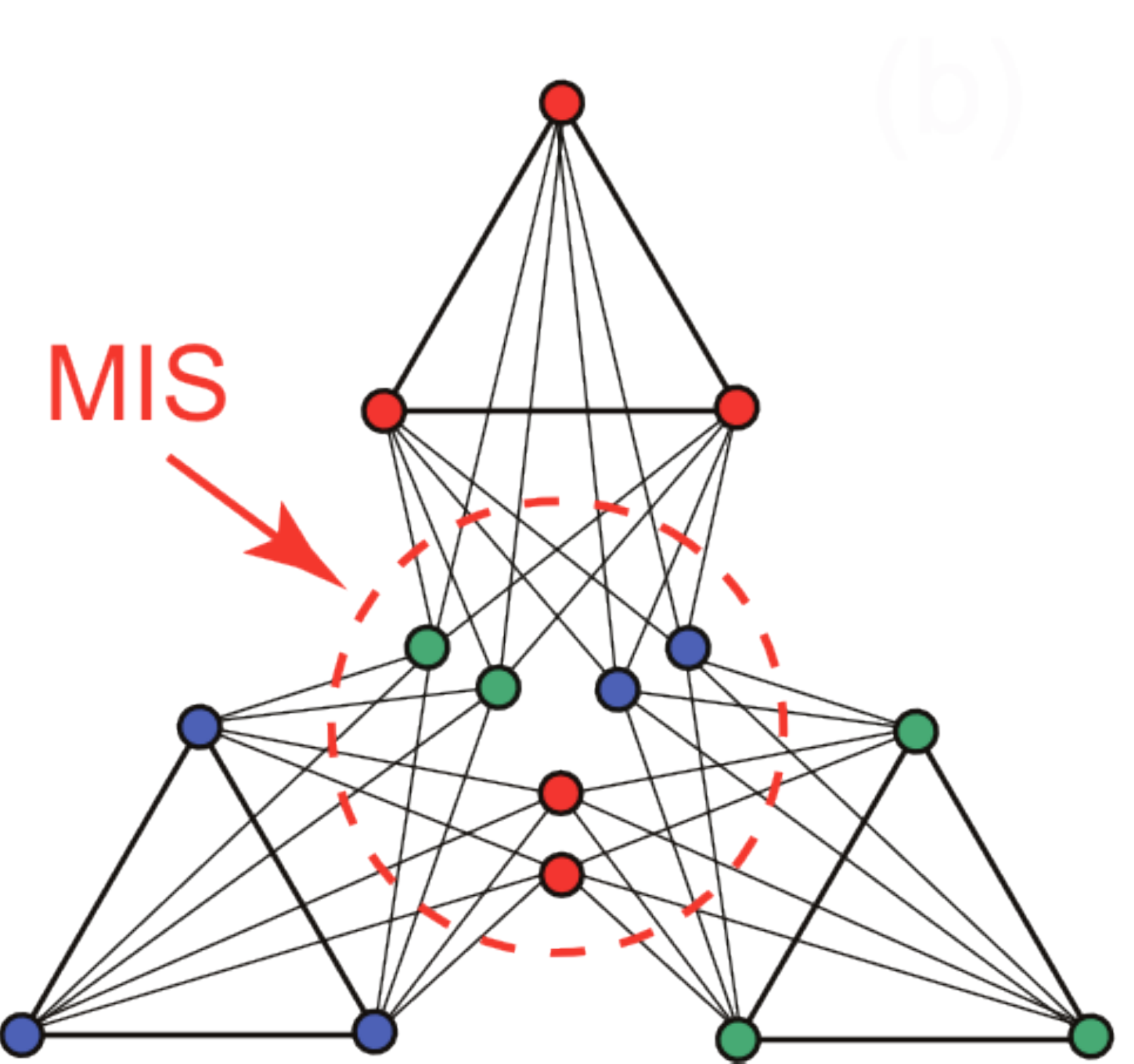}    \\
(a) & (b)
  \end{array}
$$
  \fcaption{(a) A CK graph for $r=3$ and $g=3$. The graph consists
  of 15 vertices:  $V_A=\{1,\ldots, 6\}$  forms an
    independent set of size 6, while $V_B$, consisting of $g(=3)$ groups of $r(=3)$ triangles:
    $\{7,8,9\}$, $\{10,11,12\}$, and $\{13,14,15\}$, forms $3^3$
    independent sets of size 3. The graph is connected as follows. The 6
    vertices in $V_A$ are divided into 3 groups: $\{1,2\}$, $\{3,4\}$, and
    $\{5,6\}$. The vertices in each group are adjacent to vertices in
    two groups of three triangles in $V_B$ (as illustrated by
  different colors). 
(b) The drawing of the graph with explicit connections.
The weight of a vertex in $V_A$ ($V_B$ resp.) is $w_A$ ($w_B$ resp.).
We set $w_A=1$,  and consider $1 \le w_B < 2$.
For explanation purpose, we represent a vertex in $V_A$ by a $\bullet$, and a vertex in $V_B$ by a $\triangle$.
Therefore, $V_A=\{\bullet,\bullet,\bullet,\bullet,\bullet,\bullet,\}$, forms the MIS of weight $6$; while
$\{\triangle, \triangle, \triangle\}$ is a maximal independent set of weight $3w_B(<6)$. 
  }
\label{fig:G1}
\end{figure}

\noindent{\bf CK Graph Construction.}
Let $r, g$ be integers, and $w_A$, $w_B$ be positive rational
numbers. Our graphs are specified by these four parameters. 
There
are two types of vertices in the graph: vertices of a $2g$-independent
set, denoted by $V_A$, and vertices of $g$ $r$-cliques
(which form $r^g$ maximal independent sets), denoted by $V_B$.
The weight of 
vertex in $V_A$ ($V_B$ resp.) is $w_A$ ($w_B$ resp.). 
The graph is connected as follows. Partition the vertices in $V_A$ into $g$ groups of
2 (independent) vertices. There are also $g$ groups of $r$-cliques in $V_B$. We label
both groups accordingly such that each group in $V_A$ is adjacent to all but one (the same
label) $r$-cliques in $V_B$.
Note if $w_B<2w_A$,
then we have $V_A$ forming the (global) maximum independent sets of
weight $2gw_A$, while there are $r^g$ (local) maximal independent
set of weight $gw_B$. 
See Figure
\ref{fig:G1} for an example of a graph  for
$r=3$ and $g=3$.
In general, there are infinitely many such graphs specified by the
parameters $r$ and $g$.

\subsection{\desev{} for the MIS{} Adiabatic Algorithm on  a 15-vertex CK Graph}
In the section, we fix the CK graph with $r=3$, $g=3$ as illustrated in Figure~\ref{fig:G1}.
We set $w_A=1$,  and consider $1 \le w_B < 2$.
The graph $\GCK$ consists
  of 15 vertices:  $V_A=\{1,\ldots, 6\}$  forms the maximum-weight
    independent set of weight $6$; 
while $V_B$, consisting of $3$ groups of $3$ triangles:
    $\{7,8,9\}$, $\{10,11,12\}$, and $\{13,14,15\}$, forms  $3^3$ maximal 
    independent sets of weight $3w_B < 6$.

According to Eq.(\ref{eq:Ising}), the problem Hamiltonian (and thus the adiabatic algorithm)
for MIS on $\GCK$ is 
\begin{equation}
\ham_{1} = \sum_{i \in V_A} (6J -2) \sigma_i^z + \sum_{i \in V_B} (6J -2w_B) \sigma_i^z 
+ J\sum_{ij \in \edge(G)} \sigma^z_i \sigma^z_j
\label{eq:unscaled}
\end{equation}
Here we fix $J_{ij}=J=2>w_B$ for all $ij \in \edge(\GCK)$.

%To explain \desev{} on the CK graph, we introduce the following notation:
\paragraph{Notation on representing the computational states.} For  a computational state $\ket{x_1x_2 \ldots x_n}$ where $x_i \in \{0,1\}$,
we adopt the zero position representation, namely, represent it by $\ket{i_1i_2\ldots i_k}$ where $x_{j}=0$ if and only if $j=i_t$ for some $t$.
That is, we  represent 
$\ket{000000111111111}$ (the solution state) by  $\ket{123456}$. 
Further,  we  use a $\bullet$ to denote a vertex in $V_A$, a $\triangle$ for
a vertex in $V_B$. 
That is, the solution state is now represented
by $\ket{\bullet \bullet \bullet \bullet \bullet \bullet}$,
while $\ket{\triangle \triangle \triangle}$, corresponding to a local maximal independent set of weight $3w_B$ with one vertex from each triangle.

\paragraph{Maximum vs Minimum.} The maximum of MIS corresponds to the minimum of the Ising energy. 
For explanation purpose, instead of referring to the  energy values of the Ising Hamiltonian, we will refer  to
 the values of MIS given by the pseudo-boolean function $\oy$ in Eq.(\ref{eq:Y}) by ``(-)energy'', where ``(-)'' is to    indicate the reverse ordering.

\paragraph{Example.} The (-)energy of $\ket{\bullet \bullet \bullet \bullet \bullet \bullet}$ is 6; while $\ket{\triangle \triangle \triangle\!\!\!-\!\!\triangle}$ is $4w_B - J$, where $\triangle\!\!-\!\!\triangle$ represents two connected vertices from $V_B$, e.g. vertex 7 and 8 in Figure~\ref{fig:G1}.
%% \begin{tabular}{ll}
%%   state & (-)energy\\
%%  $\ket{\bullet \bullet \bullet \bullet \bullet \bullet}$ & 6\\
%% $\ket{\triangle \triangle \triangle}$ & $3w_B$\\
%%  $\ket{\triangle \triangle \triangle\!\!\!-\!\!\!\triangle}$ & $4w_B - J$
%% \end{tabular}

\begin{figure}[h]
\begin{center}
 \includegraphics[width=0.8\textwidth]{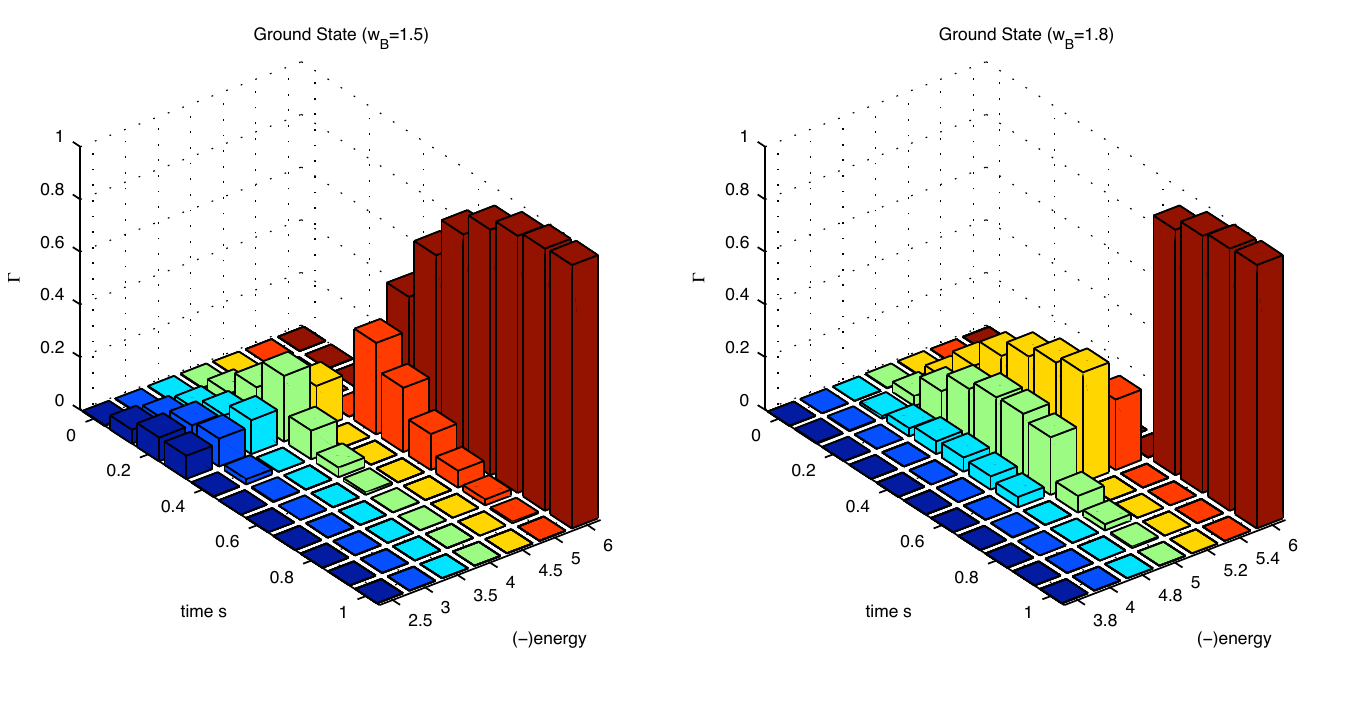}
 $s^*=0.3805, \gmin = 2.04\times10^{-2}$ \hspace*{1cm} $s^*=0.6276, \gmin = 1.04\times10^{-5}$ 
\begin{tabular}{ll||ll}
(-)energy & state & (-)energy & state\\
6 & $\ket{\bullet \bullet \bullet \bullet \bullet \bullet}$ &6 & $\ket{\bullet \bullet \bullet \bullet \bullet \bullet}$\\
5 & $\ket{\bullet \bullet \bullet \bullet \bullet}$ & 5.4 & $\ket{\triangle \triangle \triangle}$\\
4.5 & $\ket{\triangle \triangle \triangle}$ & 5.2 & $\ket{\triangle \triangle \triangle\!\!\!-\!\!\!\triangle}$\\
4 & $\ket{\bullet \bullet \bullet \bullet}$ & 5 & $\ket{\bullet \bullet \bullet \bullet \bullet}$ +  $\ket{\triangle \triangle\!\!\!-\!\!\!\triangle \triangle\!\!\!-\!\!\!\triangle}$\\
3.5 & $\ket{\bullet \bullet \triangle}$ & 4.8 &$\ket{\triangle\!\!\!-\!\!\!\triangle \triangle\!\!\!-\!\!\!\triangle \triangle\!\!\!-\!\!\!\triangle}$\\
3 & $\ket{\bullet \bullet \bullet}$ & 4 & $\ket{\bullet \bullet \bullet \bullet}$\\
2.5 & $\ket{\bullet \triangle}$ & 3.8  & $\ket{\bullet \bullet \triangle}$
\end{tabular}

\end{center}
 \fcaption{\desev{} (only the 7 lowest energy levels shown) of the ground state of the MIS adiabatic algorithm with $\ham_{1}$ in Eq.\ref{eq:unscaled} as the problem Hamiltonian for
 $w_B=1.5$ (left) and $w_B=1.8$ (right).
The x-axis is the time $s$. The y-axis is the (-)energy level.
 Each color corresponds to an energy level. 
The correspondence between (-)energy levels and the states are shown.
The z-axis is $\Gamma$.
$s^*$ is the position of the minimum spectral gap($\gmin$).
As time $s$ increases, one can see how $\Gamma$ of each energy level  evolves to get some sense of the evolution.
For example, for $w_B=1.5$ (left), 
for the (-)energy level 6 (which corresponds to the solution state), shown in brown, $\Gamma$ changes from almost 0 at $s=0.2$, to more than $0.4$ at $s=0.4$, 
to almost $1.0$ at $s=0.8$.
For $w_B=1.8$ (right), $\Gamma$ of (-) energy level 6 changes from almost $0$ before $s=0.6$ to more than $0.9$ at $s=0.7$;
while
$\Gamma$ of (-) energy level 5.4, which corresponds to the local minima, gradually increases from $s=0$ to $0.6$,
but almost $0$ after $s=0.6$.
}
  \label{fig:gsd}
\end{figure}

See Figure~\ref{fig:gsd} for the \desev{} of the 
the ground state of the adiabatic algorithm with $\ham_{1}$ in Eq.(\ref{eq:unscaled}) as the problem Hamiltonian
for $w_B=1.5$ and $1.8$.

%\paragraph{The lowest level energies.} 
%In this case, the $\wmis = 6$, corresponding to $\wmis =\{1,2,3,4,5,6\}$  which corresponds to the minimum energy 
%of the problem Hamiltonian, we denote it by (-) energy. Note each of the 27 local maximum independent sets 
%corresponds to the energy of (-) $3*w_B$. For $w_B=1.5$, we have $4.5$, while it is $5.4$ for $w_B=1.8$.
%\vnote{need to revise...}

\subsection{FQPT and Perturbation Estimation}
%\paragraph{Numerical Results on Spectral Gaps and First Order Phase Transition}}
To gain better understanding, in \cite{AC09}, we vary the weights of vertices: fix
$w_A = 1$, while varying $w_B$ from $1$ to $1.9$ with a step size of $0.1$. That is, we
fix the global maximum independent set, while increasing the weight of
the local
maximum. 
As the weight of $w_B$ increases, 
%the transition from high
%energy distribution to lower energy distribution in the instantaneous
%ground state gets more abrupt, compare Figure~\ref{fig:ed} (a) \& (d).
%And therefore, we can expect 
the minimum spectral gaps get smaller and
smaller (indeed, from $10^{-1}$ to $10^{-8}$ as $w_B$ changes from $1$ to
$1.9$ as shown in Table~\ref{table1}). 
\begin{table}[h]
    \begin{center}
\begin{tabular}{|l|c|c|}
\hline
$w_B$ & $s^*$ & $\gmin$ \\ \hline
1.0&	0.2368&	5.23e-01\\ \hline
1.1&	0.2517&	4.12e-01\\ \hline 
1.2&	0.2708&	2.90e-01\\ \hline
1.3&	0.2964&	1.68e-01\\ \hline
1.4&	0.3323&	7.14e-02\\ \hline
1.5&	0.3805&	2.04e-02\\ \hline
1.6&	0.4422&	3.63e-03\\ \hline
1.7&	0.5217&	3.39e-04\\ \hline
1.8&	0.6276&	1.04e-05\\ \hline
1.9&	0.7758&	4.14e-08\\ \hline
%\fcaption{$w_B$ changes from $1$ to $1.9$}
\end{tabular}
    \end{center}
\fcaption{The minimum spectral gap $\gmin$ (and position $s^*$) changes as $w_B$ changes from $1$ to $1.9$, for the (unscaled) problem Hamiltonian $\ham_1$ in Eq.(\ref{eq:unscaled}).}
\label{table1}
\end{table}

  \begin{figure}
$$
\begin{array}{ccl}
& (\mbox{Zoom:} s=0.627 \dots 0.628) &\\
\includegraphics[width=0.33\textwidth]{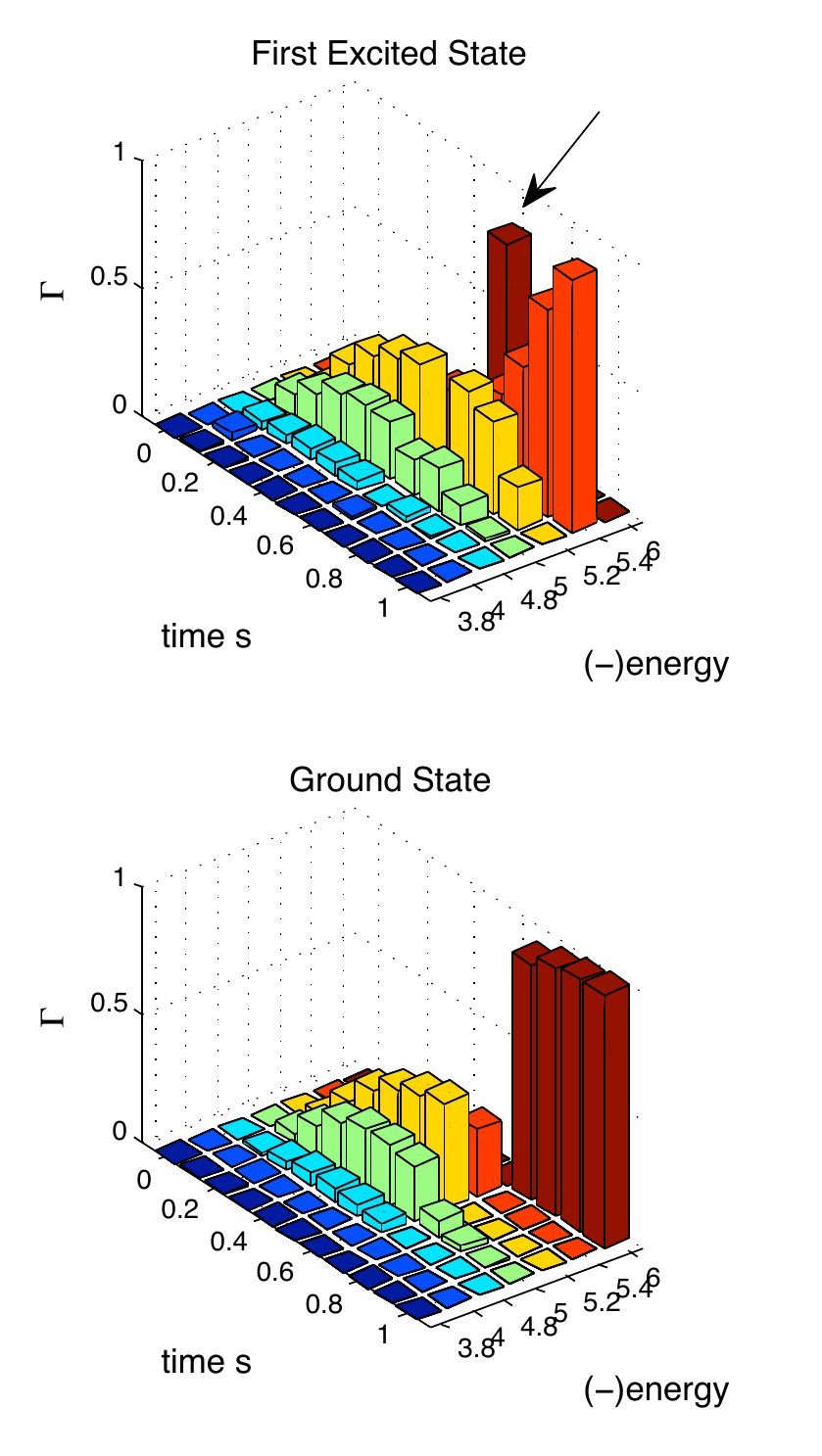}                
&
\includegraphics[width=0.3\textwidth]{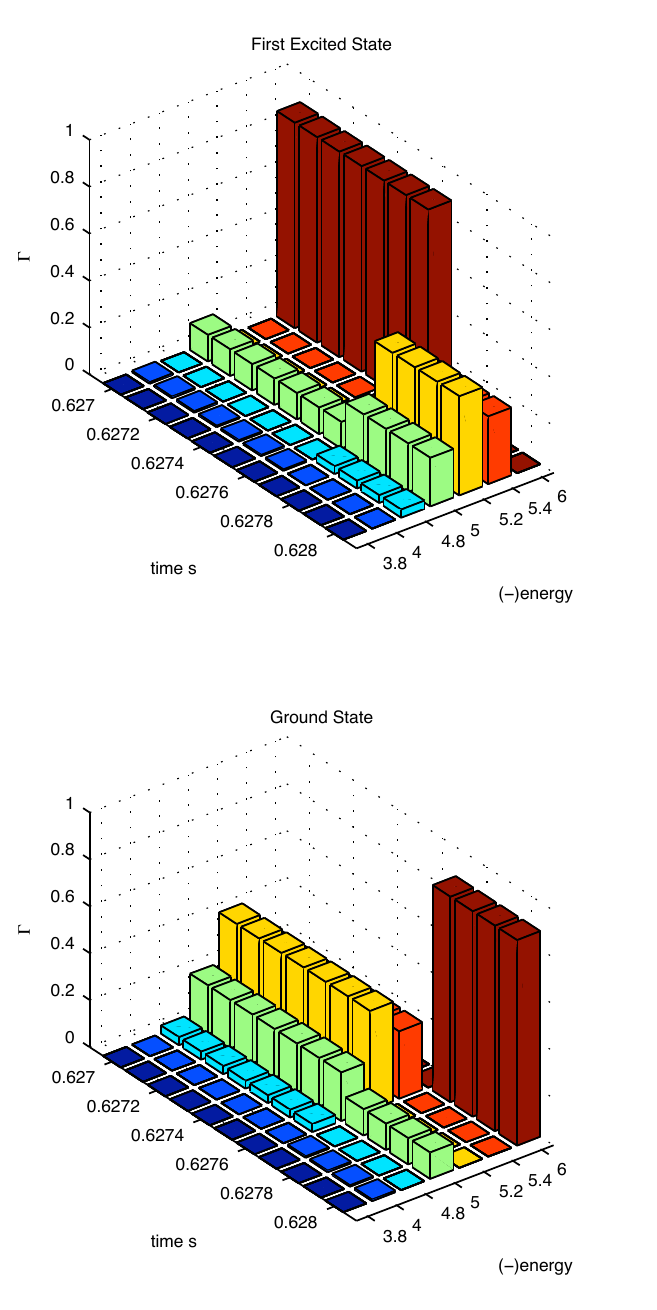}                
&\includegraphics[width=0.3\textwidth]{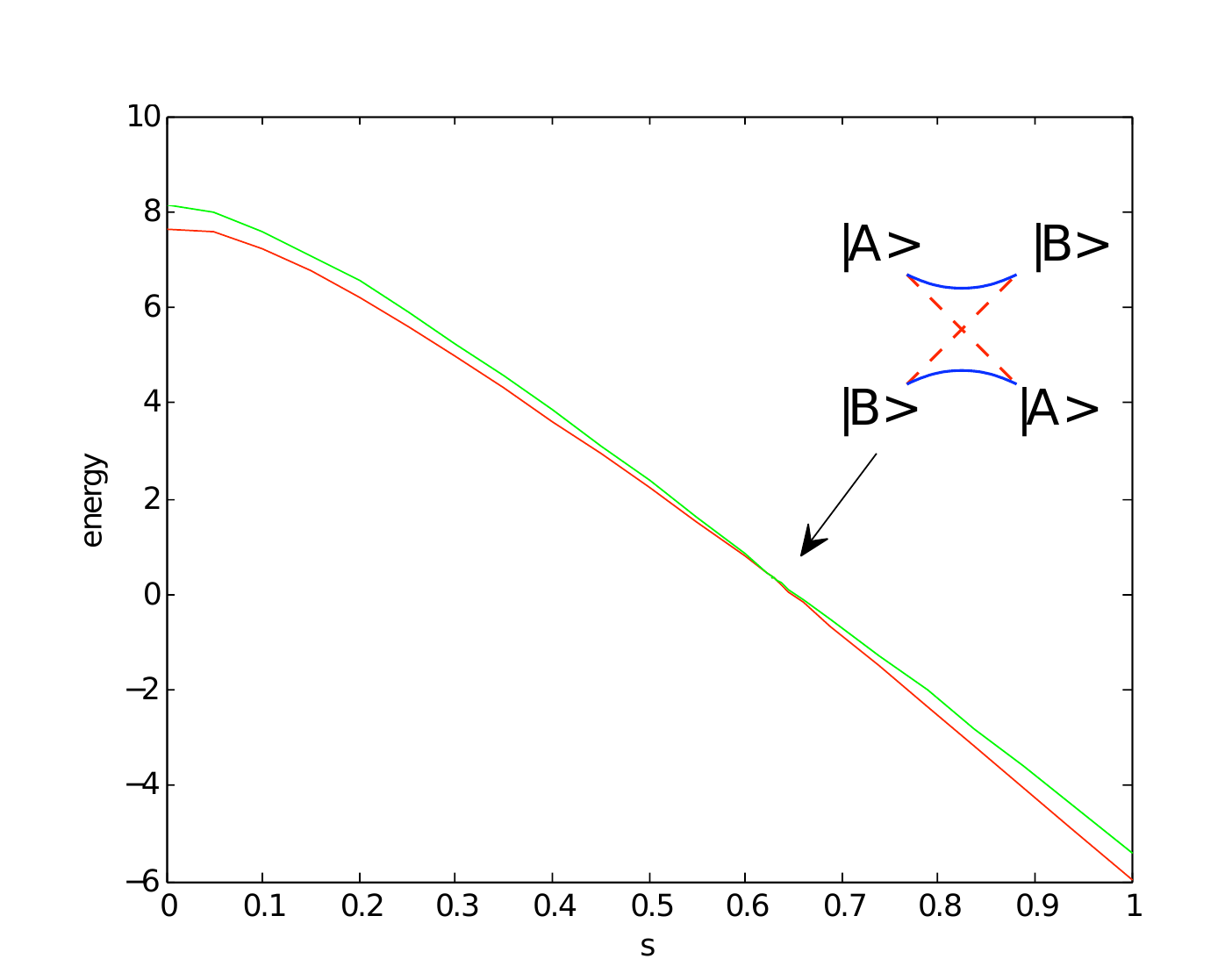}   \\
s^*=0.6276, \gmin = 1.04\times10^{-5}           & &\\
(a) & (b) & (c)
\end{array}
$$
{\tiny
  \begin{tabular}{llllllll}
     & 3.8 & 4 & 4.8 & 5 & 5.2 & 5.4& 6 \\
& $\ket{\bullet \bullet \triangle}$ & $\ket{\bullet \bullet \bullet \bullet}$ & $\ket{\triangle\!\!\!-\!\!\!\triangle \triangle\!\!\!-\!\!\!\triangle \triangle\!\!\!-\!\!\!\triangle}$ & $\ket{\bullet \bullet \bullet \bullet \bullet}$ +  $\ket{\triangle \triangle\!\!\!-\!\!\!\triangle \triangle\!\!\!-\!\!\!\triangle}$ & $\ket{\triangle \triangle \triangle\!\!\!-\!\!\!\triangle}$ & $\ket{\triangle \triangle \triangle}$ &
 $\ket{\bullet \bullet \bullet \bullet \bullet \bullet}$\\ 
 \end{tabular}
}

  \fcaption{\desev{} of the ground state and the first excited state of the MIS adiabatic algorithm with $\ham_{1}$ in \ref{eq:unscaled} as the problem Hamiltonian
for $w_B=1.8$  (a) $s=0 \ldots 1$; (b) Zoom in $s=0.627 \ldots 0.628$; (c) The lowest two energy levels of $\ham(s)$, $s=0 \ldots 1$. 
The inset illustrates a level anti-crossing between two states $\ket{B}$ and $\ket{A}$, or the system has a FQPT from $\ket{B}$ to $\ket{A}$ at the anti-crossing $s^*$. 
%This means that at time $s^{-}$ ($=s^*-\epsilon$, for some ``sufficiently'' small $\epsilon$)
% the ground state $\ket{E_0(s)}$ of the system is ``predominant'' by 
%$\ket{B}$ while the first excited state $\ket{E_1(s)}$ is ``predominant'' by $\ket{A}$. At the anit-crossing $s^*$, the two states interchange. That means, $\ket{E_0(s^+)} \sim \ket{A}$ and $\ket{E_1(s^+)} \sim \ket{B}$, where $s^+$ ($=s^*+\epsilon$).  In this case,  the spectral gap at $s^*$ (=$E_1(s^*) - E_0(s^*)$) can be approximated by the tunneling
%amplitude between  $\ket{A}$ and $\ket{B}$.
%Here we say
%the system has a FQPT or an level-anticrossing at $s^*$ where   
%$\ket{A}$ is the global minimum $\ket{\bullet \bullet \bullet \bullet \bullet \bullet}$, and 
% $\ket{B}$ is the superposition of the local minima $\ket{\triangle \triangle \triangle}$.
%More precisely, 
%this is only an ``approximate'' FQPT, where 
In this example,
$\ket{A} = \ket{\bullet \bullet \bullet \bullet \bullet \bullet}
+ \ket{\bullet \bullet \bullet \bullet \bullet}$ and $\ket{B} = \ket{\triangle \triangle\!\!\!-\!\!\!\triangle \triangle\!\!\!-\!\!\!\triangle} + \ket{\triangle \triangle \triangle\!\!\!-\!\!\!\triangle} + \ket{\triangle \triangle \triangle}$.
%(Here we explain FQPT and level anti-crossing based on the physics intuition described in \cite{AC09,altshuler-2009}.  The more precise definitions will require a quantitative specification of ``predominant'' and the value of $\epsilon$.)
}
  \label{fig:desev1}
\end{figure}

This was consequently explained by the FQPT in \cite{AC09}. 
By FQPT,  here we mean that there is a level anti-crossing between two states as illustrated in 
Figure~\ref{fig:desev1}.
The minimum spectral gap ($\gmin$) and the position ($s^*$)
were then estimated  based on the assumption of the level anti-crossing between the global minimum
and the local minima using perturbation method.
In particular, $\gmin$ was estimated by the tunneling amplitude between the global minimum and the local minima.
The formula so derived involves combinatorial enumeration of the all possible paths between local
minima  and the global minimum, and suggested $\gmin$ is exponentially (in terms of the problem size) small.
%See Figure~\ref{fig:desev1} (zoom illustrates the first order phase transition), and also the explanation by Altshuler~et~al.~\cite{altshuler-2009}.
 See also \cite{altshuler-2009,AC09,farhi-2009, young-2009} for more explanation on the FQPT and the level anti-crossing.
%\vnote{need revision...}
%\vnote{need to draw a figure to illustrate...}

\section{Varying Parameters in the Problem Hamiltonian for MIS}
\label{sec:change-parameter}
In this section, we show that by changing the parameters in the
problem Hamiltonian for MIS on the CK graph, 
the FQPT no longer occurs and 
we can significantly increase $\gmin$.

%\vnote{perhaps better wording...}.
% and the first order phase transition no longer occurs.

%% Recall that according to Theorem \cite{}, the QUBO function for an instance $<G,\{c_i\}>$ is
%% $$ \oy(x_1,\ldots, x_n) = \sum_{i \in \ver(G)}c_i x_i - \sum_{ij \in \edge(G)}
%%   J_{ij}x_ix_j$$
%% where $J_{ij}$ is required at least $\min\{c_i,c_j\}$, for each $ij \in \edge{G}$,
%% and the corresponding problem Hamiltonian is
%% \begin{equation}
%% \ham_{F_0} = \sum_{i \in \ver(G)} h_i \sigma^z_i + \sum_{ij \in \edge(G)} J_{ij}
%% \sigma^z_i \sigma^z_j
%% \end{equation}
%% where $h_i = \sum_{j \in \nbr(i)}
%%   J_{ij} - 2c_i$, for $i \in \ver(G)$.

Recall that in the pseudo-boolean formulation of MIS as in Theorem~\ref{thm:mis}, 
 the requirement for $J_{ij}$ is at least $\min\{c_i,c_j\}$, for each $ij \in \edge(G)$.
For simplicity,  we consider the simplest case in which $J_{ij}=J$ for all $ij \in \edge(G)$.
In other words, we have the corresponding problem Hamiltonian:
\begin{equation}
\ham_{1} = \sum_{i \in \ver(G)} (d_iJ-2c_i) \sigma^z_i + \sum_{ij \in \edge(G)} J
\sigma^z_i \sigma^z_j\\
%= J (\sum_{i \in \ver(G)} (d_i-2c_i/J) \sigma^z_i + \sum_{ij \in \edge(G)} 
%\sigma^z_i \sigma^z_j)
\end{equation}
where $d_i$ is the degree of vertex $i \in \ver(G)$.

The natural question is: how does the \art{} change when we vary $J$? Note that it is not sufficient to
consider only the minimum spectral gap change (as almost all the other works on adiabatic quantum computation did)
because by increasing $J$, the maximum energy of the system Hamiltonian also increases.
Instead, in order to keep the maximum energy of the system Hamiltonian comparable, we keep $J$ fixed and
vary $c_i$  instead, namely multiplying all weights $c_i$ by a scaling factor, say  $1/k$, for $k \ge 1$, 
which does not change the original
problem to be solved. We remark that this is equivalent to multiplying $J$ by $k$, and then multiply the problem Hamiltonian by ($1/k$). 
%The details and more general case can be found in \cite{Precision-scale}.

That is, we consider the following 
(scaled) problem Hamiltonian 
\begin{eqnarray}
\ham_k= \sum_{i \in \ver(G)} (Jd_i-2c_i/k )\sigma^z_i + \sum_{ij \in \edge(G)} J 
\sigma^z_i \sigma^z_j
\label{eq:Hk}
\end{eqnarray}
where $k \ge 1$ is the scaling factor.

\subsection{Minimum Spectral Gap $\gmin$ Without FQPT}
\begin{figure}[h]
$$
\begin{array}{cc}
k=1 & k=10 \\
\includegraphics[width=0.45\textwidth]{K1_annotate.pdf}                
&
\includegraphics[width=0.4\textwidth]{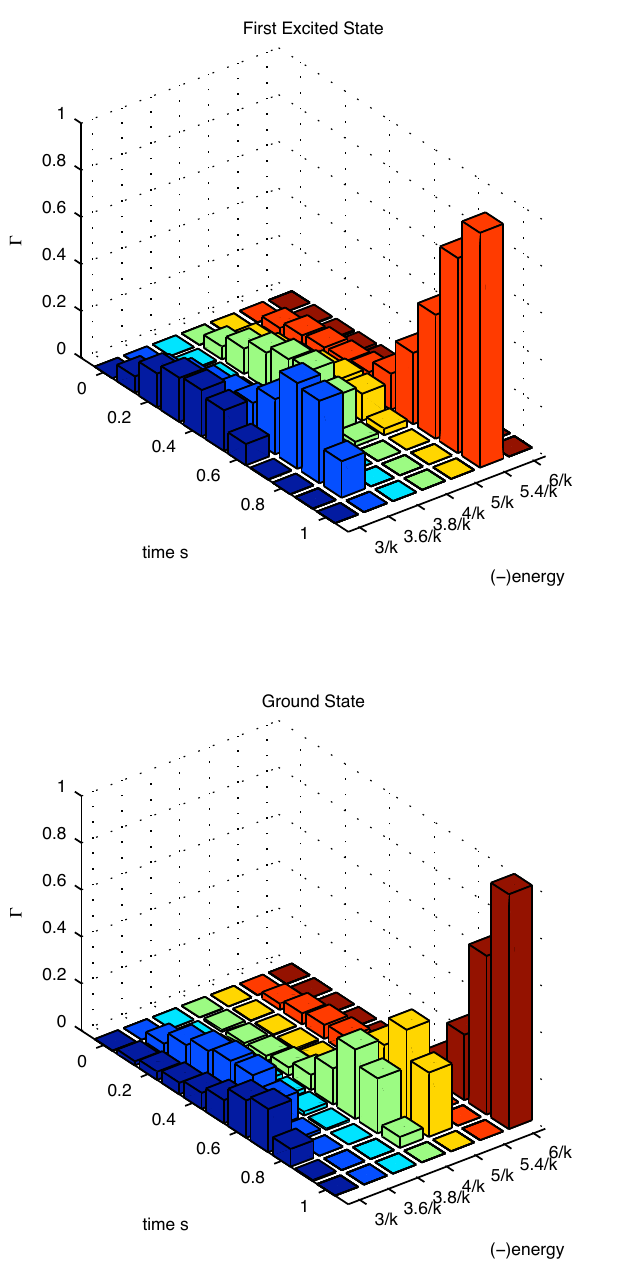} \\
s^* = 0.627637, \gmin=1.04\times10^{-5} & s^*=0.667731, \gmin=0.145
\end{array}
$$

{\tiny
  \begin{tabular}{llllllll}
    $k=1$ & 3.8 & 4 & 4.8 & 5 & 5.2 & 5.4& 6 \\
& $\ket{\bullet \bullet \triangle}$ & $\ket{\bullet \bullet \bullet \bullet}$ & $\ket{\triangle\!\!\!-\!\!\!\triangle \triangle\!\!\!-\!\!\!\triangle \triangle\!\!\!-\!\!\!\triangle}$ & $\ket{\bullet \bullet \bullet \bullet \bullet}$ +  $\ket{\triangle \triangle\!\!\!-\!\!\!\triangle \triangle\!\!\!-\!\!\!\triangle}$ & $\ket{\triangle \triangle \triangle\!\!\!-\!\!\!\triangle}$ & $\ket{\triangle \triangle \triangle}$ &
 $\ket{\bullet \bullet \bullet \bullet \bullet \bullet}$\\ \\
$k=10$ & $3/k$ & $3.6/k$ & $3.8/k$ & $4/k$ & $5/k$ & $5.4/k$ & $6/k$\\
& $\ket{\bullet \bullet \bullet}$ & $\ket{\triangle \triangle}$ & $\ket{\bullet \bullet \triangle}$ &
$\ket{\bullet \bullet \bullet \bullet}$ & $\ket{\bullet \bullet \bullet \bullet \bullet}$ & $\ket{\triangle \triangle \triangle}$
& $\ket{\bullet \bullet \bullet \bullet \bullet \bullet}$
 \end{tabular}
}
\fcaption{\desev{} of the ground state and the first excited state of the MIS adiabatic algorithm with  problem Hamiltonian $\ham_{1}$ (left)
% \ref{eq:unscaled} as the problem Hamiltonian
and  $\ham_{10}$ (right)
where $w_B=1.8$. 
%The x-axis is the time $s$; the y-axis is the energy level.
% Each color corresponds to an energy level.
Notice the differences in the lowest few excited states.
For $k=1$, the $2nd$ and $3rd$ excited states are  superpositions of $\triangle$s (vertex in $V_B$ which constitutes the local optima);
while for $k=10$, the $2nd$ and $3rd$ excited states are  superpositions of $\bullet$s (vertex in $V_A$ which constitutes the global optimum). As a result, the first order phase transition from local minima to global minimum occurs for $k=1$, which results in the $\gmin= 1.04\times 10^{-5}$ at $s^*=0.627$. For $k=10$, such crossing no longer occurs, and $\gmin=0.145$ at $s^*=0.667$.  
See Figure~\ref{fig:scaled-zoom} for the zoom-in.
}
  \label{fig:scaled}
\end{figure}

\begin{figure}[h]
$$
\begin{array}{ccc}
k=1 (s:0.627 \ldots 0.628)  & k=1 (s:0.62763 \ldots 0.62764)  &k=10 (s: 0.667 \ldots 0.668)\\
\includegraphics[width=0.3\textwidth]{K1_Zoom.pdf}                
&
\includegraphics[width=0.3\textwidth]{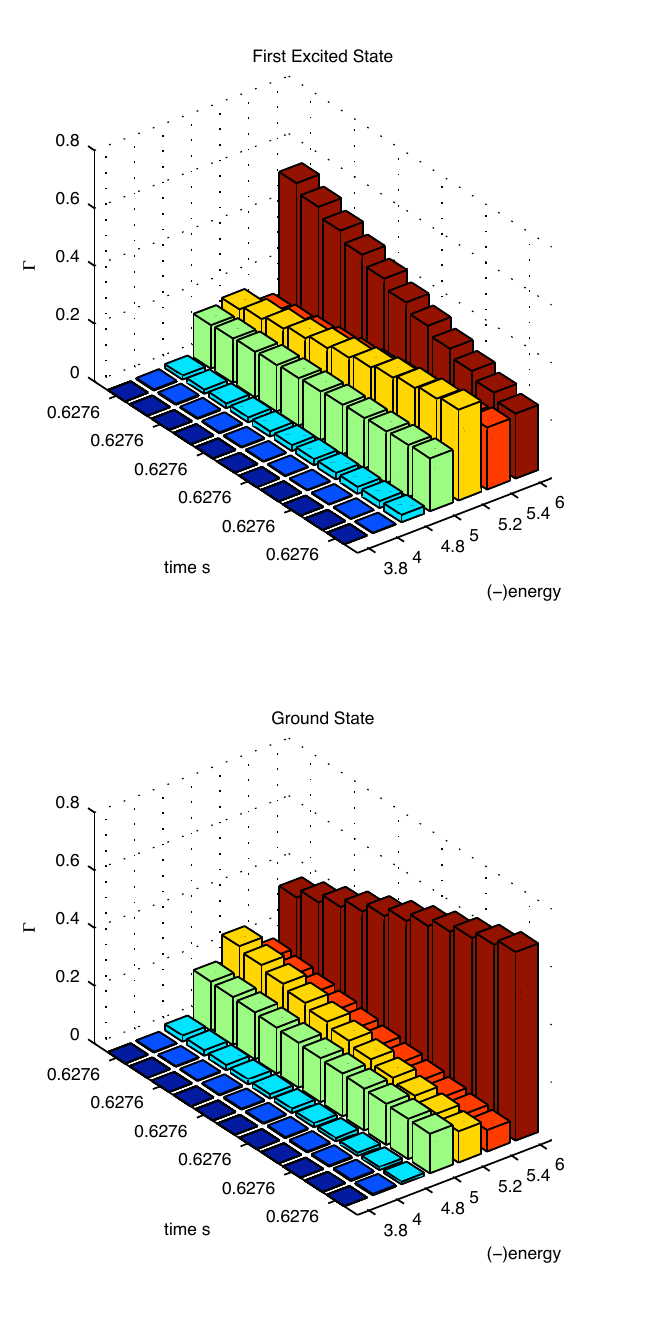}                
&
\includegraphics[width=0.3\textwidth]{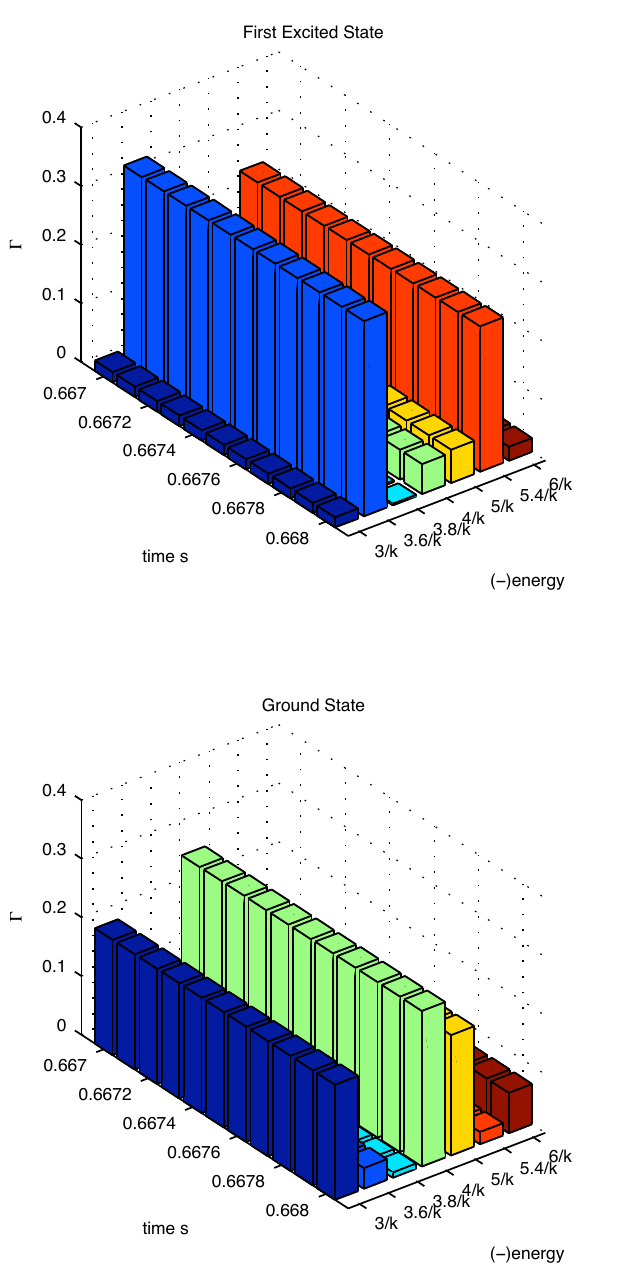}                \\
s^*= 0.6276, \gmin=1.04\times 10^{-5} & s^*= 0.6276, \gmin=1.04\times 10^{-5}  & s^*=0.6677, \gmin=1.45 \times 10^{-1}\\

\end{array}
 $$
{\tiny
  \begin{tabular}{llllllll}
    $k=1$ & 3.8 & 4 & 4.8 & 5 & 5.2 & 5.4& 6 \\
& $\ket{\bullet \bullet \triangle}$ & $\ket{\bullet \bullet \bullet \bullet}$ & $\ket{\triangle\!\!\!-\!\!\!\triangle \triangle\!\!\!-\!\!\!\triangle \triangle\!\!\!-\!\!\!\triangle}$ & $\ket{\bullet \bullet \bullet \bullet \bullet}$ +  $\ket{\triangle \triangle\!\!\!-\!\!\!\triangle \triangle\!\!\!-\!\!\!\triangle}$ & $\ket{\triangle \triangle \triangle\!\!\!-\!\!\!\triangle}$ & $\ket{\triangle \triangle \triangle}$ &
 $\ket{\bullet \bullet \bullet \bullet \bullet \bullet}$\\ \\
$k=10 $ & $3/k$ & $3.6/k$ & $3.8/k$ & $4/k$ & $5/k$ & $5.4/k$ & $6/k$\\
& $\ket{\bullet \bullet \bullet}$ & $\ket{\triangle \triangle}$ & $\ket{\bullet \bullet \triangle}$ &
$\ket{\bullet \bullet \bullet \bullet}$ & $\ket{\bullet \bullet \bullet \bullet \bullet}$ & $\ket{\triangle \triangle \triangle}$
& $\ket{\bullet \bullet \bullet \bullet \bullet \bullet}$
 \end{tabular}
}

  \fcaption{Zoom around the position $s^*$ of the minimum spectral gap. }
%For $k=1$, there is a FQPT around $s^*$.  No such transition for $k=10$.}
  \label{fig:scaled-zoom}
\end{figure}

\begin{figure}
$$
\begin{array}{ccc}
k=1 & k=2 &k=3\\
\includegraphics[width=0.3\textwidth]{K1_annotate.pdf} & \includegraphics[width=0.25\textwidth]{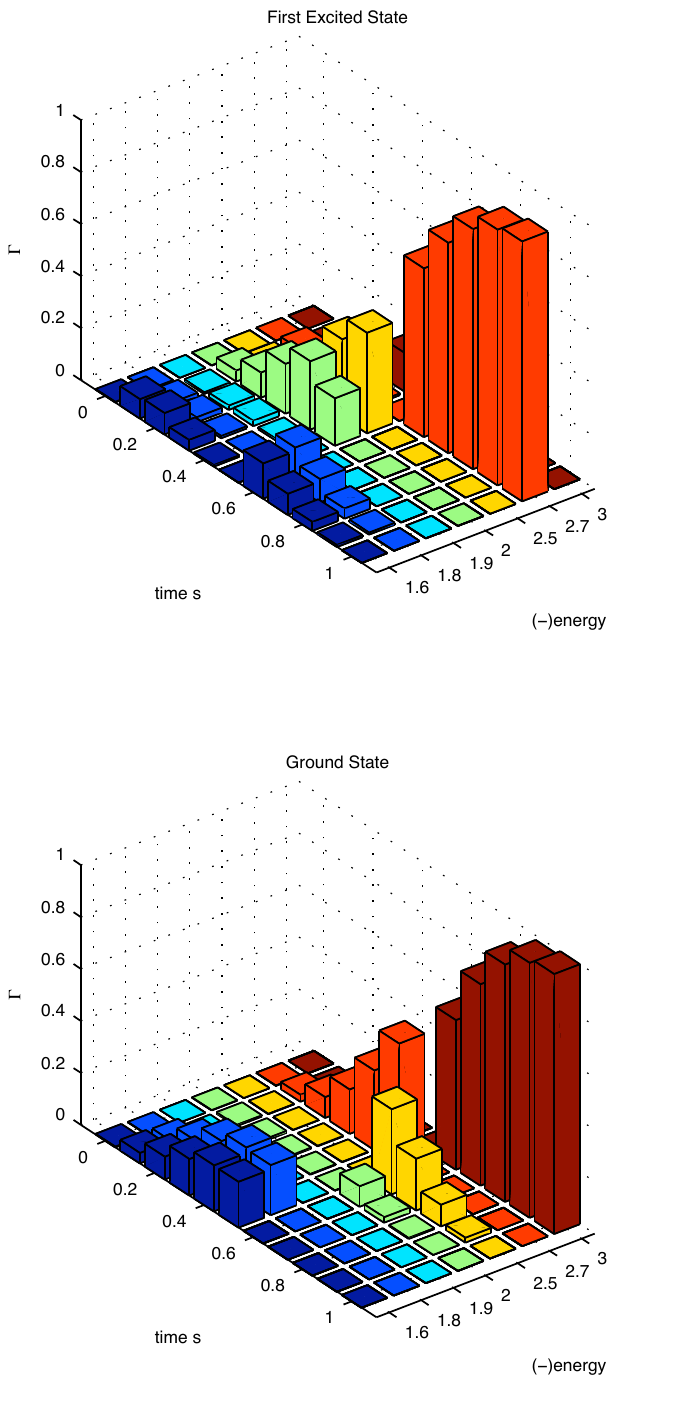} & \includegraphics[width=0.25\textwidth]{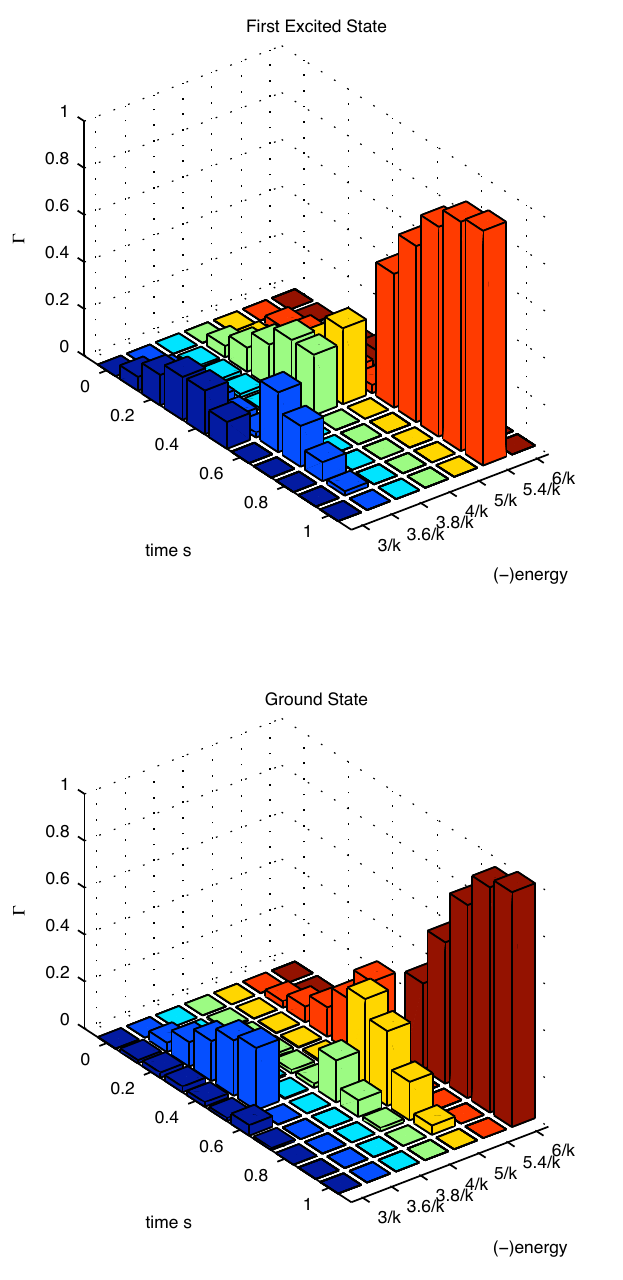}\\
s^*= 0.6276, \gmin=1.04\times 10^{-5} & s^*=0.5457,	\gmin=6.37\times 10^{-3}  & s^*=0.5446,	\gmin=3.30\times10^{-2}\\
k=5 & k=10 & k=50\\
\includegraphics[width=0.25\textwidth]{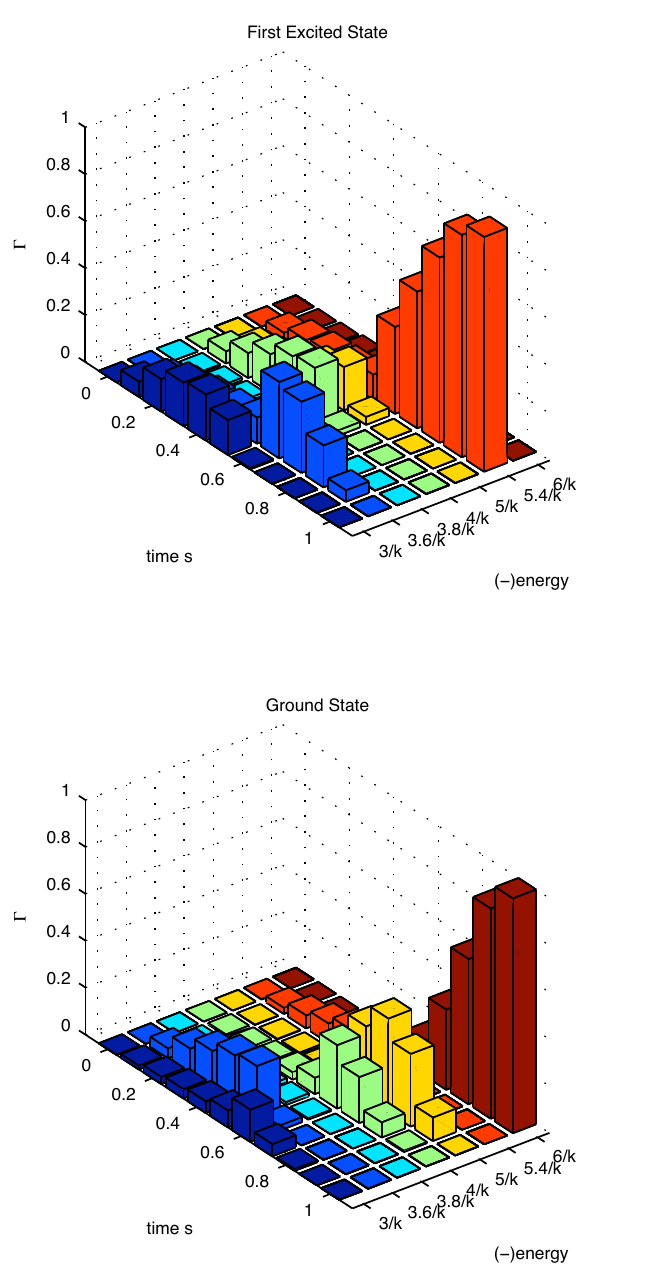} & \includegraphics[width=0.25\textwidth]{K10.pdf} & \includegraphics[width=0.25\textwidth]{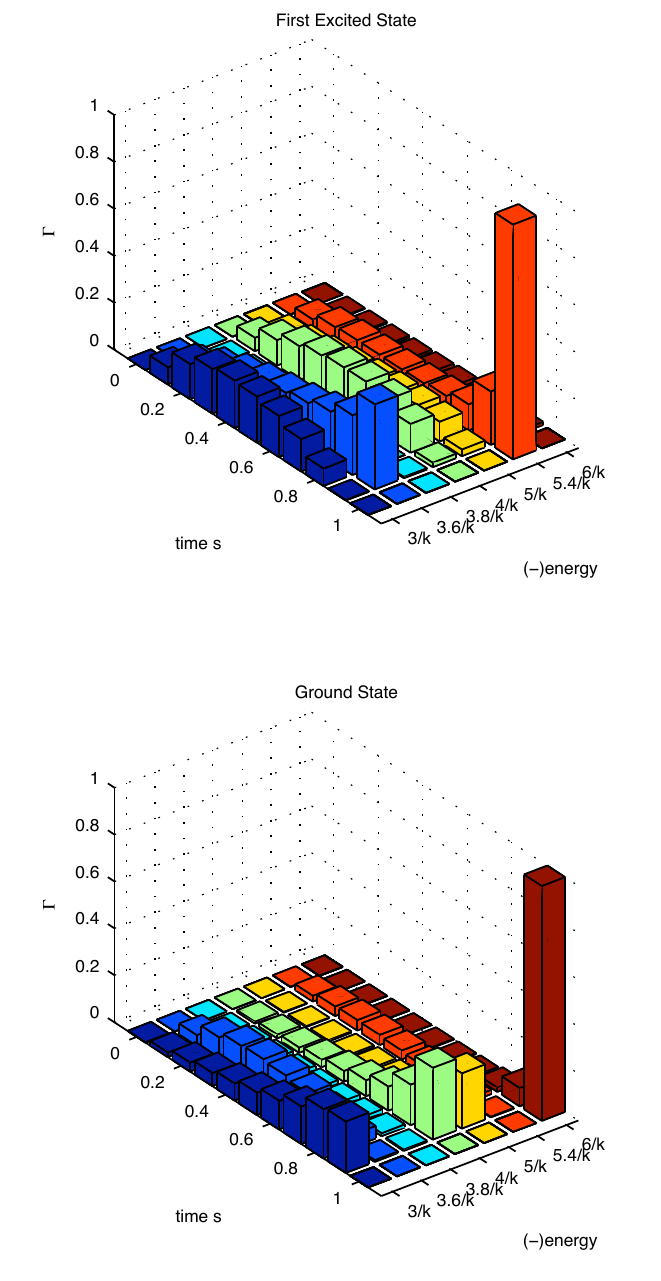}\\
s^*=0.5741, \gmin=9.67\times 10^{-2} & s^*=0.6677, \gmin=1.45 \times 10^{-1} & s^*=0.9977,\gmin=4.79 \times 10^{-2}
\end{array}
$$ 
{\tiny
  \begin{tabular}{llllllll}
    $k=1$ & 3.8 & 4 & 4.8 & 5 & 5.2 & 5.4& 6 \\
& $\ket{\bullet \bullet \triangle}$ & $\ket{\bullet \bullet \bullet \bullet}$ & $\ket{\triangle\!\!\!-\!\!\!\triangle \triangle\!\!\!-\!\!\!\triangle \triangle\!\!\!-\!\!\!\triangle}$ & $\ket{\bullet \bullet \bullet \bullet \bullet}$ +  $\ket{\triangle \triangle\!\!\!-\!\!\!\triangle \triangle\!\!\!-\!\!\!\triangle}$ & $\ket{\triangle \triangle \triangle\!\!\!-\!\!\!\triangle}$ & $\ket{\triangle \triangle \triangle}$ &
 $\ket{\bullet \bullet \bullet \bullet \bullet \bullet}$\\ \\
$k=2 $ & $1.6$ & $1.8$ & $1.9$ & $2$ & $2.5$ & $2.7$ & $3$\\
& $\ket{\triangle \triangle\!\!\!-\!\!\!\triangle}$ & $\ket{\triangle \triangle}$ & $\ket{\bullet \bullet \triangle}$ &
$\ket{\bullet \bullet \bullet \bullet}$ & $\ket{\bullet \bullet \bullet \bullet \bullet}$ & $\ket{\triangle \triangle \triangle}$
& $\ket{\bullet \bullet \bullet \bullet \bullet \bullet}$\\ \\

$k\ge3 $ & $3/k$ & $3.6/k$ & $3.8/k$ & $4/k$ & $5/k$ & $5.4/k$ & $6/k$\\
& $\ket{\bullet \bullet \bullet}$ & $\ket{\triangle \triangle}$ & $\ket{\bullet \bullet \triangle}$ &
$\ket{\bullet \bullet \bullet \bullet}$ & $\ket{\bullet \bullet \bullet \bullet \bullet}$ & $\ket{\triangle \triangle \triangle}$
& $\ket{\bullet \bullet \bullet \bullet \bullet \bullet}$
 \end{tabular}
}

 \fcaption{\desev{} of the ground state and the first excited state of the adiabatic algorithm with  problem Hamiltonian
$\ham_k$ for $w_B=1.8$, where $k=1,2,3,5,10,50$.
$s^*$ is the position of the minimum spectral gap($\gmin$).
}
  \label{fig:different-K}
\end{figure}

%\paragraph{Estimation of $\gmin$.} 
The \desev{}s of $\ham_{1}$ and $\ham_{10}$  are shown in Figure~\ref{fig:scaled} and Figure~\ref{fig:scaled-zoom}.
The anti-crossing between the global minimum and the local minima (for $k=1$) 
no longer occurs for $k=10$, and $\gmin$ increases from
$1.04\times10^{-5}$ to $1.45\times 10^{-1}$. 
A worthwhile observation is 
the change in the lowest few excited energy levels: 
 for $k=1$, the lowest few excited states (beyond the first excited state) 
of the problem Hamiltonian  is mainly the superposition of states from $V_B$ ($\triangle$) (which constitutes the local minima); while 
these states of the scaled ($k=10$) problem Hamiltonian is mainly the superposition of states from $V_A$($\bullet$) (which constitutes the global minimum).
That is, the minimum spectral gap can be increased drastically (by as
much as four order of magnitude in this example) when the second or higher excited energy levels are changed (while the lowest and first excited energy level stay the same).
The \desev{}s of $\ham_k$ for $k=1,2,3,5,10,50$ are shown in Figure~\ref{fig:different-K}.
%From the Figure, we see that the FQPT does not occur for $k \ge 2$. 
%$\gmin$ increases as $k$ increases from $1$ to $10$, but decreases from $10$ to $50$.

In \cite{AC09}, 
based on the FQPT assumption, 
we estimate $\gmin$ (for $\ham_{1}$) by the tunneling amplitude between the local minima and the global minimum,
which suggests that $\gmin$ is exponentially small. 
However, for $k=10$,
from our numerical data and \desev{} in Figure~\ref{fig:scaled}, we see that the FQPT 
(that causes $\gmin$ to be exponentially small)  
no longer 
occurs, and $\gmin$ increases significantly.
This seems to suggest that $\gmin$ to be polynomially small instead
for a general CK graph of size $n$.
The problem for analytically estimating $\gmin$
 of $\ham_{k}$ for a general CK graph of size $n$ remains open.
We remark here that the 
perturbation method is still valid (in fact, 
 as we increase $k$, we also increase the minimum spectral gap position $s^* \tends 1$), however we can no longer assume that $\gmin$ can be approximated by the tunneling amplitude between the two (localized) states.
%, but instead by two mixtures of states.

%% See Figure~\ref{fig:different-K} for the \desev  of the ground state and the first excited state of the MIS adiabatic algorithm for $\ham_k$ with $k=1,2,3,5,10,50$.
%% The numerical data for different $k$s are shown in Table~\ref{table2}.
%% Observe that as we increase $k$ from 1 to 10,  $\gmin$ increases, however from $20$ to $50$ $\gmin$ decreases.
%% This perhaps can be explained by that as $k$ increases, the final energy difference also decreases, and it becomes dominant when $k$ is large enough.
%% So how does $\art$ change with $k$?

\subsection{Scaling Factor and \art{}}
\label{sec:art}
In this section, we discuss what  the good scaling factor should be, and how it affects the \art.
To address this question, we need an appropriate formulation for \art.
Notice that even for numerical studies, it is not sufficient to just
consider $\gmin$ (as the other numerical works on adiabatic
quantum computation did, see e.g. \cite{young-2009}), but the matrix element of the time 
derivative of the Hamiltonian also matters.
In particular, we adopt the following three formulations,
which are related to the widely used traditional condition:
$$
(*)
\left\{
\begin{array}{l}
  \art_1(\ham) = \frac{\max_{0 \le s \le 1}\mat(s)}{\gmin^2}\max_{0 \le s \le 1}||\ham(s)||  \\
  \art_2(\ham) = \frac{\mat(s^*)}{\gmin^2}\max_{0 \le s \le 1}||\ham(s)||, \mbox{ where } \gmin=E_1(s^*) - E_0(s^*)\\
  \art_3(\ham) = \max_{0 \le s \le 1}\frac{\mat(s)}{(E_1(s) - E_0(s))^2}\max_{0 \le s \le 1}||\ham(s)|| 
\end{array}
\right.
$$
where $\mat(s)=|\bra{E_1(s)}\frac{d\ham}{ds}\ket{E_0(s)}|$ is the matrix element of the time derivative Hamiltonian at time $s$,
and $\ham(s) \ket{E_i(s)} = E_i(s)\ket{E_i(s)}$.
See Table~\ref{table2} for the numerical comparisons.

    \begin{table}[h]
%      \begin{center}
	\begin{tabular}{|l|l|l|l|l|l|l|l|}
\hline
$k$ & $s^*$ & $\gmin $ & $\mat(s^*)$ &  $\max_{s}\mat(s)$   & $\max_{s}||\ham||$ & $\art_2$ & $\art_1$\\
\hline
1  &	0.62763727 &	1.04e-05 &	4.02e+00 &	4.02e+00 &	2.26e+02 &	8.34e+12 &		8.34e+12 \\ \hline 
2  &	0.54578285 &	6.37e-03 &	2.04e+00 &	2.04e+00 &	2.48e+02 &	1.24e+07 &		1.24e+07 \\ \hline 
3  &	0.54467568 &	3.30e-02 &	1.41e+00 &	1.41e+00 &	2.55e+02 &	3.32e+05 &		3.32e+05 \\ \hline 
4  &	0.55610853 &	6.83e-02 &	1.18e+00 &	1.18e+00 &	2.59e+02 &	6.57e+04 &		6.58e+04 \\ \hline 
5  &	0.57419149 &	9.67e-02 &	1.06e+00 &	1.07e+00 &	2.61e+02 &	2.96e+04 &		2.99e+04 \\ \hline 
10 &	0.66773072 &	1.45e-01 &	7.48e-01 &	7.92e-01 &	2.66e+02 &	9.45e+03 &		1.00e+04 \\ \hline 
20 &	0.80170240 &	1.30e-01 &	4.72e-01 &	5.68e-01 &	2.68e+02 &	7.48e+03 &		9.01e+03 \\ \hline 
30 &	0.99318624 &	7.97e-02 &	8.95e-09 &	4.26e-01 &	2.69e+02 &	3.78e-04 &		1.80e+04 \\ \hline 
40 &	0.99642154 &	5.99e-02 &	4.90e-10 &	4.35e-01 &	2.69e+02 &	3.67e-05 &		3.26e+04 \\ \hline 
50 &	0.99779592 &	4.79e-02 &	5.30e-11 &	4.41e-01 &	2.69e+02 &	6.20e-06 &		5.16e+04 \\ \hline 
	\end{tabular}

\vspace*{0.5cm}

\begin{tabular}{|l|l|l|l|l|l|l|}
\hline
$k$ & $s'$ & $g(s') $ & $\mat(s')$ & $\frac{\mat(s')}{g(s')^2}$    & $\max_{s}||\ham||$ & $\art_3$ \\
\hline
1 &	0.62763727 &	1.04e-05 &	4.02e+00 &	3.70e+10 &	2.26e+02 &	8.34e+12 \\ \hline 
2 &	0.54578226 &	6.37e-03	&     2.04e+00	&     5.02e+04	&     2.48e+02	&     1.24e+07 \\ \hline
3 &	0.54461081 &	3.30e-02 &	1.41e+00 &	1.30e+03 &	2.55e+02 &	3.32e+05 \\ \hline 
4 &	0.55545411 &	6.83e-02 &	1.18e+00 &	2.54e+02 &	2.59e+02 &	6.57e+04 \\ \hline 
5 &	0.57223394 &	9.68e-02 &	1.07e+00 &	1.14e+02 &	2.61e+02 &	2.97e+04 \\ \hline 
10&	0.65682886 &	1.46e-01 &	7.75e-01 &	3.64e+01 &	2.66e+02 &	9.66e+03 \\ \hline 
20&	0.77115481 &	1.33e-01 &	5.41e-01 &	3.08e+01 &	2.68e+02 &	8.24e+03 \\ \hline 
30&	0.83962780 &	1.08e-01 &	4.43e-01 &	3.82e+01 &	2.69e+02 &	1.02e+04 \\ \hline 
40&	0.88050519 &	8.82e-02 &	3.93e-01 &	5.05e+01 &	2.69e+02 &	1.36e+04 \\ \hline 
50&	0.90581875 &	7.39e-02 &	3.63e-01 &	6.64e+01 &	2.69e+02 &	1.79e+04 \\ \hline 
	\end{tabular}

%      \end{center}
where $g(s) = E_1(s) - E_0(s)$, and $s' = \argmax_{0 \le s \le 1}\frac{\mat(s)}{g(s)^2}$. 

\fcaption{$\art_1$, $\art_2$, $\art_3$ for $\ham_k$  in Eq.(\ref{eq:Hk}).
Observations: (1)  $\gmin$ increases as $k$ increases  from 1 to 10, but decreases from $10$ to $50$. 
(2) $\art_1$, $\art_2$, and $\art_3$ are close for $k<5$.
(3) The matrix element $\mat(s^*)$ at the position of minimum spectral gap is extremely small for $k \ge 30$.
(4) For $k>10$, $s^*$ (the position of the minimum spectral gap) 
is different from $s'$, where $s' = \argmax_{0 \le s \le 1}\frac{\mat(s)}{g(s)^2}$.  
}
\label{table2}
    \end{table}

%From Table\ref{table2}, we have the following observations.
%\paragraph{Observations.} 

%For $k=10$, $\gmin$ increases significantly. This might be explained by the absence of the FQPT,  which causes the exponential
%small gap for $k=1$.
From Table \ref{table2}, we see that $\gmin$ increases as $k$ increases  from 1 to 10, however, decreases from $10$ to $50$ (even though it is still much larger than $k=1$).
The latter, perhaps,  can be explained by the following: 
 as $k$ increases, the difference between the low energy levels decreases, and becomes dominate for $k>10$.
%So how do we determine the best value for $k$? 
We remark that the optimal value for $k$ seems to depend only on  the vertex weights (for which $J$ depends on),
and  independ of the problem size.
By increasing the scaling factor, we also increase the 
 precision (or dynamic range) requirement for representing the parameters ($h_i$ \& $J_{ij}$) in the problem Hamiltonian, 
which is one of the important physical resources.

Notice that $\art_2 \le \art_1 \le \art_3$. The condition given in $\art_3$ is the formula that one would derive
 from the adiabatic approximation. The condition in $\art_1$ is the widely used traditional version. The condition
in $\art_2$ was mentioned in \cite{Young}. The natural question is: when are they asymptotically equivalent?  
Indeed, 
the three versions of \art{}  coincide for some Hamiltonians (e.g. for $k=1$). 
However, they can be very different for the large $k$. The main reason is that the matrix element $\mat(s)$ can be extremely small at the minimum spectral gap position $s^*$. For example, for $k=50$, $s^* \tends 1$, $\mat(s^*)$ is extremely small. Note one can show that $\mat(s)=|\bra{E_1(s)} \ham_{\ms{init}}
  \ket{E_0(s}|/s$ for $s\in (0,1]$ because $ \mat(s)= |\bra{E_1(s)}\ham(1) - \ham(0)\ket{E_0(s}| = |\bra{E_1(s)}\frac{\ham(s) - \ham(0)}{s}\ket{E_0(s}| =|\bra{E_1(s)}\ham(0)\ket{E_0(s}|/s$. 
  Thus, for our initial Hamiltonian, $\mat(s)$ measures the overlap of the states with one single bit flip, and in this case it is extremely small.
Observe that
the position of the minimum spectral gap $s^*$ is not the same as the position $s'$ where $\frac{\mat(s)}{g(s)^2}$ is maximized.
What should be the appropriate formulation of $\art$? Should it be $\art_3$? If so, under what condition, can $\art_1$ be a good approximation to $\art_3$?
and under what condition, can we assume that $\gmin$ is the dominating factor (as have been assumed by all other works)? 
%Here we raise these questions, and perhaps the researchers in adiabatic theorem can directly address these questions.

\section{Discussion}
\label{sec:discussion}

In this paper, 
we have shown that by 
changing the parameters in the problem Hamiltonian (without changing the problem to be solved) of the adiabatic algorithm for 
MIS on the CK graph, we prevent the FQPT, that causes the exponential small $\gmin$, from occurring and significantly increase $\gmin$.
We do so by scaling the vertex-weight of the graph, namely, multiplying the weights of vertices by a scaling factor.
In order to determine the best scaling factor, we raise the basic question about what the appropriate formulation of adiabatic running time should be.

%\paragraph{Quantum vs classical local search.}
In \cite{DMV01,DV01}, van Dam et al.~argued that adiabatic quantum optimization might be thought of as a kind of ``quantum local search'', 
 and  in \cite{DV01}, they constructed a special family of 3SAT instances for which the clause-violation cost function based adiabatic algorithm 
required exponential time.
% Farhi~et~al.~\cite{diff-path1} showed that the exponential small gap
% could be overcome by different initial Hamiltonians. 
In \cite{ChoiDiff}, we point out that the exponential small gap argument does not apply to a different adiabatic algorithm for 3SAT.
Our CK graph was designed to trap local search algorithms
in the sense that there are many local minima to mislead the local search process.
From \desev{} on a 15-vertex CK graph,
 we see that indeed this is the case for $\ham_1$ and the adiabatic algorithm would require exponential time due to the
exponential small $\gmin$ caused by  the FQPT or the level anti-crossing between the global minimum and the local minima.
However, for $\ham_k$ (say $k=10$), the FQPT no longer occurs and $\gmin$ increases significantly, which might suggest the possibility of exponential speed-up
over $\ham_1$.
It remains challenging on how to analytically bound $\gmin$ and/or $\art$ of the adiabatic algorithm for $\ham_{k}$ on general
(CK) graphs. One worthwhile observation from this work is that the
minimum spectral gap can be increased drastically when the second or
higher excited energy states are overlapping with the ground state in
spite of the large amount of first excited states (which constitutes
the local minima). Recall that NP-complete problems can be polynomial
reducible to each other. The reduction requires only the solution  to be preserved,
i.e. there is 
 a polynomial time algorithm that maps
the solution to the 
reduced problem  to  the solution to the original problem and vice versa.
In other words, the reduction might only preserves the solution
(i.e. the ground state) and alter the energy levels of the problem
Hamiltonian. Therefore, according to the observation, different reduction is possible to give rise to different problem Hamiltonians,
and thus different adiabatic quantum  algorithms, for the same problem. Indeed, we
have shown in \cite{ChoiDiff} that 
based on the NP-complete reductions, 
we describe different adiabatic quantum algorithms to which 
the arguments in \cite{DV01,altshuler-2009} for the failure of their adiabatic
quantum algorithms do not apply.

In summary, although our result is only numerical and supported by visualization,
this small example, nevertheless, serves to  show that it is possible to avoid FQPT,
and also to concretely
clarify that
it is not sufficient to consider one specific problem Hamiltonian
(and thus one specific adiabatic quantum optimization algorithm) for
proving  the failure of adiabatic quantum optimization for a problem.
%  as in
%\cite{DV01,altshuler-2009}.

%For our small size, our scaled problem Hamiltonian seem to suggest polynomial gaps.
%We are currently investigating on how to derive the gaps analytically.

% The design and analysis of the continuous adiabatic algorithm remain challenging.

\nonumsection{Acknowledgements}
\noindent
I would like to thank my very enthusiastic students in my adiabatic quantum computing class: 
Ryan Blace,
Russell Brasser, 
Mark Everline,
    Eric Franklin,
     Nabil Al Ramli,
  and   Aiman Shabsigh, who also helped to name \desev.
I would like to thank Siyuan Han, Peter Young, Mohammad Amin, Neil Dickson, Robert Raussendorf, Tzu-Chieh Wei and Pradeep Kiran for their comments. 
Thanks also go to David Sankoff and David Kirkpatrick for the encouragement.

%\newpage
%\centerline{Appendix}

\nonumsection{References}

\end{document}